\documentclass[reprint,superscriptaddress,amsmath,amssymb,aps,prb,twocolumn, nobibnotes]{revtex4-2}

\usepackage{graphicx}
\usepackage{dcolumn}
\usepackage{bm}
\usepackage{color}
\usepackage{breqn}

\usepackage{setspace}
\usepackage{ragged2e}

\begin{document}

\title[phonon thermoreflectance]{Infrared Phonon Thermoreflectance in Polar Dielectrics}

\affiliation{$Department~of~Mechanical~and~Aerospace~Engineering,~University~of~Virginia,~Charlottesville,~Virginia~22904, USA$}

\affiliation{$Department~of~Materials~Science~and~Engineering,~University~of~Virginia,~Charlottesville,~Virginia~22904, USA$}
\affiliation{$Department~of~Physics,~University~of~Virginia,~Charlottesville,~Virginia~22904, USA$}

\author{Saman Zare}
\thanks{These two authors contributed equally.}
\author{William D. Hutchins}
\thanks{These two authors contributed equally.}
\author{Daniel Hirt}
\author{Elizabeth Golightly}

\affiliation{$Department~of~Mechanical~and~Aerospace~Engineering,~University~of~Virginia,~Charlottesville,~Virginia~22904, USA$}

\author{Patrick E. Hopkins}
\email{Corresponding Author: phopkins@virginia.edu}
\affiliation{$Department~of~Mechanical~and~Aerospace~Engineering,~University~of~Virginia,~Charlottesville,~Virginia~22904, USA$}
\affiliation{$Department~of~Materials~Science~and~Engineering,~University~of~Virginia,~Charlottesville,~Virginia~22904, USA$}
\affiliation{$Department~of~Physics,~University~of~Virginia,~Charlottesville,~Virginia~22904, USA$}

\begin{abstract}
In this work, we investigate dielectric materials for thermoreflectance-based thermal sensing by extracting key optical parameters using temperature-dependent spectroscopic ellipsometry in the mid-infrared regime. Leveraging optical phonon resonances, we demonstrate that the thermoreflectance coefficients in polar dielectrics rival, and in some cases exceed by an order of magnitude, those observed in commonly used metals that are typically used as temperature transducers in thermoreflectance measurements. We introduce a transducer figure of merit (FOM) that combines pump absorption and probe reflectance modulation at different wavelengths, serving as a design-oriented screening metric for comparing thermoreflectance transducer performance across materials and spectral regions. Our results show that polar materials can exhibit performance up to eight times greater than that of metal transducers. To demonstrate practical capability, we perform transient
thermoreflectance measurements on a 100 nm
thermally grown SiO$_2$ film on silicon.  These results position dielectric materials as compelling candidates for next-generation thermal metrology, broadening the design space for optical thermometry, with strong implications for high-resolution thermal mapping and characterization of layered device structures based on phonon probing.

\end{abstract}

\maketitle
\onecolumngrid

\twocolumngrid

Modulation spectroscopy techniques, such as thermoreflectance, piezoreflectance, and electroreflectance~\cite{willardsonModulationTechniques1972}, have been fundamental in probing material systems. In thermal sciences, these methods measure periodic reflectivity perturbations induced by light absorption, forming the foundation of optical pump-probe techniques like time-domain thermoreflectance (TDTR)~\cite{cahillAnalysisHeatFlow2004}, frequency-domain thermoreflectance (FDTR)~\cite{schmidtCharacterizationThinMetal2010} and steady-state thermoreflectance (SSTR)~\cite{braunSteadystateThermoreflectanceMethod2019}. The non-contact and adaptable nature of these techniques has enabled the study of thermal transport across a wide array of material systems, from meso- to nano-scales. These techniques have advanced understanding of electron dynamics~\cite{grafNodalQuasiparticleMeltdown2011,tzallasExtremeultravioletPumpProbe2011}, electron-phonon coupling~\cite{hirtIncreasedThermalConductivity2024b, choiThermalSpintransferTorque2015b},  phonon transport~\cite{gambettaRealtimeObservationNonlinear2006,hoqueExperimentalObservationBallistic2024, zhangUltrafastRelaxationLattice2023, kennesTransientSuperconductivityElectronic2017} , and thermal management in electronic devices~\cite{aryanaSuppressedElectronicContribution2021a}.  Optical approaches are also suited for in-situ growth characterization and device quality assurance~\cite{kuzmenkoUniversalOpticalConductance2008, jamgotchianSituObservationInterferometric2001}. 

A key aspect of these techniques is a thin surface film that absorbs pump light and serves as a thermometer to characterize the resulting temperature rise with the probe. This ``transducer" is central to thermoreflectance design. Understanding its physical and chemical properties, including thermal conductivity, stability, specific heat, adhesion, and dielectric function, is essential for accurate measurements. Thereby, extensive research has focused on transducer  optimization~\cite{otterTemperaturabhaengigkeitOptischenKonstanten1961, scoulerTemperatureModulatedReflectanceGold1967, roseiThermomodulationSpectraAu1972, colavitaThermoreflectanceTestMo1983}. FDTR, TDTR, and SSTR all utilize metal transducers for two primary functions: converting optical excitation into broadband thermal response and acting as a reflective surface with well-characterized thermoreflectance, enabling thermal gradient assessment. Metals have become the standard for this role due to their high thermoreflectance~\cite{wilsonThermoreflectanceMetalTransducers2012b, halteOpticalResponsePeriodically2006} and strong visible-range absorption from inter-band transitions~\cite{dumkeInterbandTransitionsMaser1962,halteOpticalResponsePeriodically2006}, ensuring effective flux reception. Signal strength in thermoreflectance is directly influenced by the pump beam's absorption, which dictates heating amplitude, and the thermoreflectance magnitude, which determines sensitivity. Since most commonly used transducers are thin metal films deposited after the manufacturing process, thermoreflectance has primarily been limited to \textit{ex situ} characterization.

While qualitative assessments have measured thermoreflectance coefficient, $d R /dT$, these studies have been limited in both the material range~\cite{scoulerTemperatureModulatedReflectanceGold1967, roseiElectronicStructureBcc1980} and the spectral regimes~\cite{wilsonThermoreflectanceMetalTransducers2012b}.  In most applications, using metal transducers is intuitive given that pulsed lasers often operate in the visible to near-infrared, where metals exhibit strong electronic absorption and thermoreflectance. Their small optical penetration depth also ensures pump energy is abosrbed at the surface, simplifying analysis. Historically, the prevalence of laser wavelengths in the visible range~\cite{delfyettHighpowerUltrafastLaser1992, morgnerSubtwocyclePulsesKerrlens1999} reinforced the use of metal transducers.  However, advancements in optical parametric amplification~\cite{bridaGenerationBroadbandMidinfrared2007}, have expanded spectral access, prompting a reassessment of optimal transducer materials~\cite{hutchinsUltrafastEvanescentHeat2025,tomkoUltrafastChargeCarrier2025,tomkoLonglivedModulationPlasmonic2021}.

The reliance of metal transducers on inter-band transitions for high thermoreflectance~\cite{halteOpticalResponsePeriodically2006} inherently limits their application to specific wavelength ranges. Expanding options beyond metals could unlock new capabilities in thermal metrology, particularly in the mid-infrared regime, where phonon resonances in dielectrics may enhance sensitivity. Dielectrics also offer flexibility, as their transparency in certain spectral regimes enables thermal transduction at various depths within multilayer systems. Shifting focus from electronic absorption to optical phonons may provide improved thermoreflectance performance while offering greater versatility in probing buried interfaces and sub-surface thermal transport.

In this letter, we extend the range of viable transducers into the mid-infrared by leveraging optical phonons in dielectrics to achieve higher thermoreflectance coefficients for stronger signals and greater absorption for more intense thermal events. Using variable-angle spectroscopic ellipsometry, we measure $dR/dT$ for several  dielectric materials across the mid-infrared. Our temperature-dependent ellipsometry results provide quantitatively determined values of $dn/dT$ and $dk/dT$ for these materials, allowing for direct comparison of transducer efficiency.  To quantify the effectiveness of materials in pump-probe thermoreflectance measurements, we introduce a transducer figure of merit (FOM) that accounts for both pump absorption and probe signal strength. Notably, we observe that the thermoreflectance of dielectrics can exceed that of common metal transducers, yielding a higher FOM and highlighting the potential of dielectrics as effective alternatives for mid-infrared thermoreflectance applications. 

The principal parameter for transducing the optical response of a material to temperature is its reflectance. At normal incidence, it is given by the Fresnel equations as
\begin{equation}
    R=\frac{(n - 1)^2 +k^2}{(n + 1)^2 +k^2}
    \label{Eq:Reflect}
\end{equation}

\noindent where $n$ and $k$ represent the refractive index and the extension coefficient, respectively. Both $n$ and $k$ can be derived from optical measurements and are related to the complex dielectric function $\varepsilon = ({n} + i {k})^2$. The dielectric behavior of polar dielectrics in the infrared can be modeled using the standard Lorentz model~\cite{lorentzTheoryElectronsIts1909} as
\begin{equation}
    \varepsilon(\omega) = \varepsilon_\infty \left(1+\sum_j \frac{A_j^2}{\omega_j - \omega^2 + i\Gamma_j\omega}\right)
    \label{Eq:diel}
    \end{equation}

\noindent where $A_j$, and $\omega_j$ are respectively the amplitude and centroid frequency of the $j^{th}$ mode, and $\Gamma_{j}$ is the damping factor. Temperature-induced shifts and broadening of these modes give rise to strong variations in $n$, $k$, and consequently $R$, forming the physical basis for enhanced infrared thermoreflectance in polar dielectrics.

\begin{figure}[b]
\centering
 \includegraphics[width=0.5\textwidth]{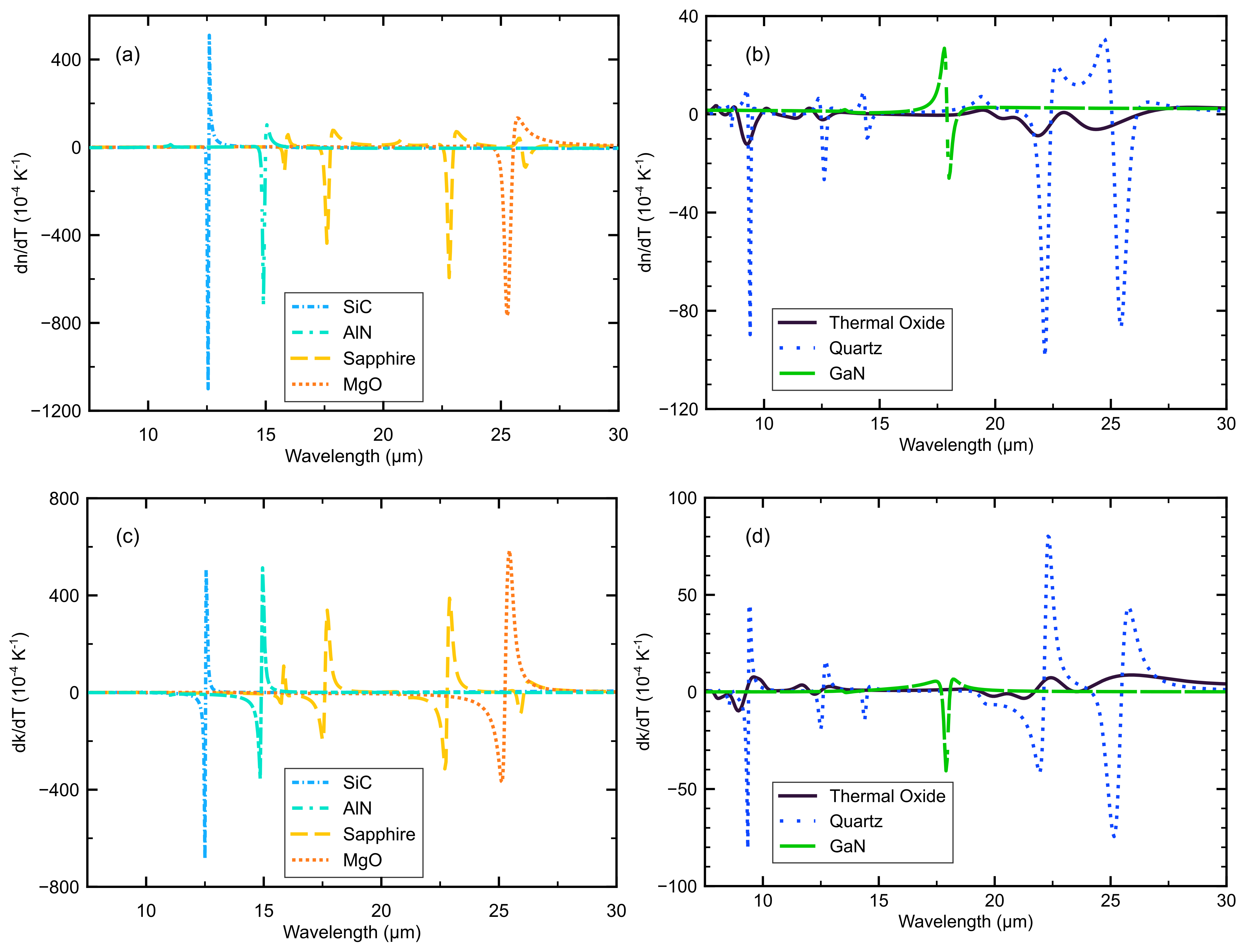}
    \caption[] {Extracted temperature dependence of the (a,b) refractive index and (c,d) extinction coefficient for SiC, AlN, sapphire, MgO, thermal oxide, Quartz, and GaN.}
  \label{fig:2_PTR}
\end{figure}

Guided by this framework, we experimentally extract thermoreflectance coefficients for a range of dielectric materials using temperature-dependent spectroscopic ellipsometry. Ellipsometry measures the change in polarization upon reflection and enables direct extraction of the wavelength-dependent optical constants. The measured sample set includes commercially available bulk samples of quartz, sapphire, silicon carbide (SiC), and magnesium oxide (MgO), along with thin films of aluminum nitride (AlN) and gallium nitride (GaN) deposited with a thickness of $\sim$2.5 $\mu$m on sapphire substrates \cite{kohBulklikeIntrinsicPhonon2020,kohHighThermalConductivity2021} along with a 100 nm thermally grown SiO$_2$ film on silicon. Sample details are provided in the Supplemental Materials. 

We perform measurements across the 2-30 $\mu$m range using a J.A. Woollam IR-VASE system. We also integrate a Linkam TSEL1000 heater stage to determine the temperature dependence of the optical properties. The thermoreflectance coefficient $dR/dT$ is computed by propagating the measured temperature-dependent refractive index and extinction coefficient through the Fresnel reflectance relation and evaluating a finite-difference derivative between 25 \textdegree C and 200 \textdegree C, consistent with prior thermoreflectance studies~\cite{wilsonThermoreflectanceMetalTransducers2012b}. This approach directly links temperature-induced changes in the optical constants to the measured reflectance modulation.

It is evident from Eq.~(\ref{Eq:Reflect}) that the refractive index plays a crucial role in defining optical behavior, leading to significant variations in transparency and reflectivity across different substrates. This transparency region can be exploited in experiment design, allowing the probe in thermoreflectance to penetrate the surface and access thermal transport at deeper interfaces. Conversely, selecting an opaque region where optical penetration depth is less than the film thickness (see Supplemental Materials) enables the technique to mimic conventional metal-transducer thermoreflectance. This flexibility underscores the importance of detailed characterization of the complex refractive index. For opaque transducer configurations, determining the exact absorption mechanism requires careful consideration.

We present the measured values for the temperature dependence of refractive index and extinction coefficient, i.e., $dn/dT$ and $dk/dT$, for our dielectric sample set in Fig. \ref{fig:2_PTR}.  A key observation is the rapid variation in $dn/dT$ and $dk/dT$ near TO phonon wavelengths, driven by temperature-induced shifts and broadenings of optical phonon resonances. At elevated temperatures, TO phonons redshift due to anharmonic effects, altering the refractive index profile. Simultaneously, increased phonon scattering at higher temperatures broadens absorption features and enhances damping. These changes collectively produce strong temperature-dependent modulation of optical properties, making spectral regions near TO phonons highly sensitive for thermoreflectance-based thermometry. Above $\sim$15 $\mu$m, changes in $n$ and $k$ with temperature become minimal, as the influence of optical phonon resonances diminishes. In this long-wavelength infrared regime, the refractive index is primarily governed by the low-frequency dielectric response, which is relatively temperature-insensitive. Consequently, materials in this range exhibit weak optical perturbations and smaller reflectivity variations under thermal excitation.

The computed thermoreflectance coefficient ($dR/dT$) for the tested dielectrics is derived from temperature-dependent complex refractive indices, as presented in Fig.~\ref{fig:3_PTR}. The results reveal high sensitivity to optical phonon resonances, with peak $dR/dT$ values occurring near TO and LO phonon frequencies. As shown in Fig. \ref{fig:3_PTR}, pronounced thermoreflectance peaks arise from strong phonon absorption modulating both the refractive index and extinction coefficient. These results confirm that phonon-driven modulations dominate thermoreflectance in these dielectrics. The high sensitivity of the refractive index to temperature variations results in a large thermoreflectance coefficient. This effect arises from both the steep dispersion of the refractive index and the rapid modulation of optical absorption, which directly influences reflectivity. Therefore, wavelengths close to TO phonon modes exhibit enhanced thermoreflectance sensitivity, making them optimal for high-precision thermometry. In addition to the TO-dominated response, elevated $dR/dT$ values are also observed near LO phonon frequencies. In this regime, the enhancement does not arise from large values of $dn/dT$ or $dk/dT$, but from the fact that at the LO phonon frequency the real part of the dielectric function, $\varepsilon_1$, approaches zero, rendering the reflectance highly sensitive to small temperature-induced shifts or broadenings of the LO resonance. Outside this region, i.e., in transparent or weakly absorbing regions, $dn/dT$ and $dk/dT$ are relatively small, leading to a lower thermoreflectance response. In this regime, reflectivity changes are primarily governed by the weakly temperature-dependent background dielectric constant, i.e. $\varepsilon_\infty$ in Eq.~\ref{Eq:diel}. This trend is consistent across all samples in the test set, reinforcing the importance of spectral selection in thermoreflectance experiment design. 

These results establish the material-level thermoreflectance response and temperature-dependent optical constants of polar dielectrics, providing key inputs for quantitative benchmarking and future studies. To demonstrate practical capability, we perform transient thermoreflectance measurements on a $\sim$100 nm thermally grown SiO$_2$ film on silicon, using the thermoreflectance response of SiO$_2$ to measure the thermal boundary conductance across the SiO$_2$/Si interface. In this measurement, shown schematically in Fig.~\ref{fig:1_PTR}a and detailed in the Supplemental Material, a green pump transmits through SiO$_2$ and is absorbed by the electrons in Si. By tuning the probe wavelength on and off the optical phonon resonance (8.8 and 6 $\mu$m, respectively), we selectively probe the oxide thermoreflectance response. As shown in Fig.~\ref{fig:1_PTR}b, the signal exhibits an early-time rise from electron excitation in silicon followed by rapid decay due to relaxation. For delay times longer than 300 ps, the on-resonance measurement shows a re-rise in thermoreflectance absent in the off-resonance data, indicating oxide heating via the heated volume in Si diffusing energy across the SiO$_2$/Si interface and into SiO$_2$. This measurement is enabled by the large thermoreflectance of SiO$_2$ at 8.8 $\mu$m. Similar behavior has been reported in TDTR measurements of bilayer metallic systems due to subsurface thermalization of pump energy~\cite{choiIndirectHeatingPt2014,giriMechanismsNonequilibriumElectronphonon2015a,tomkoLonglivedModulationPlasmonic2021}.


\onecolumngrid

\begin{figure}[b]
\centering
 \includegraphics[width=0.9\textwidth]{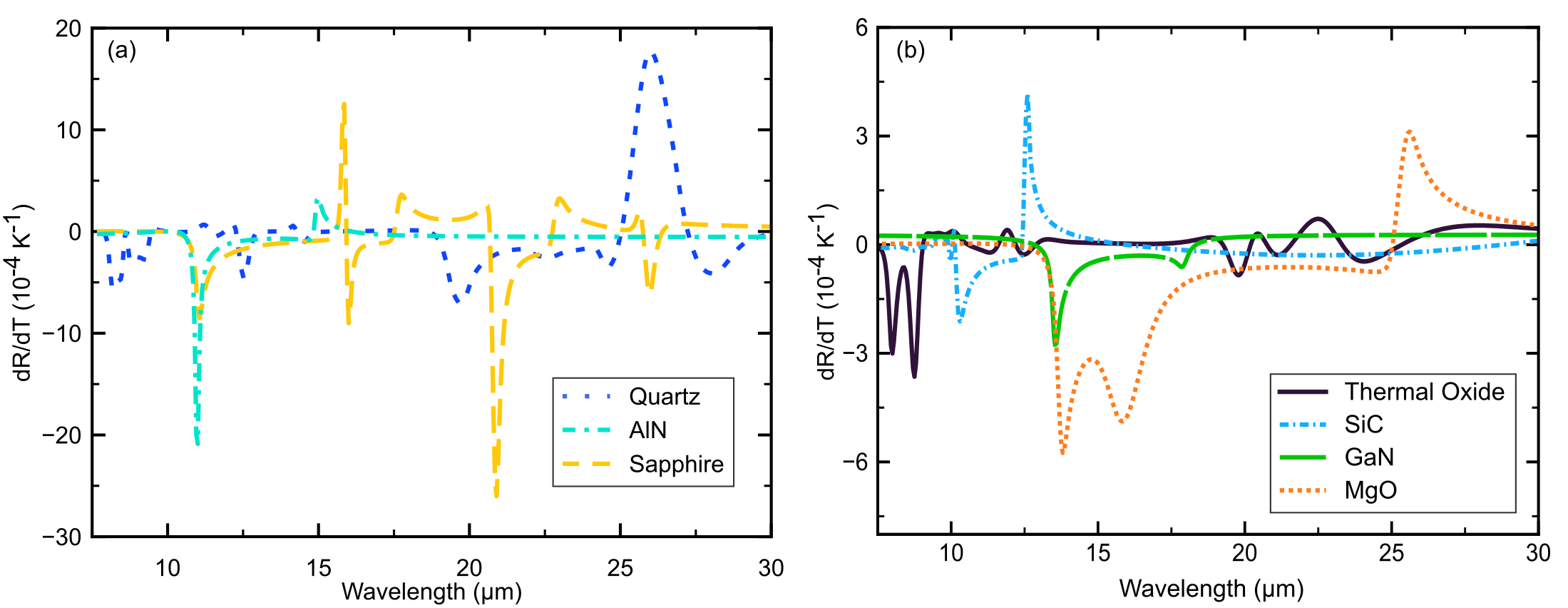}
    \caption[] {Thermoreflectance coefficients extracted from infrared spectroscopic ellipsometry for (a) quartz, AlN, sapphire, and (b) thermally  oxide, SiC, GaN, MgO.}
  \label{fig:3_PTR}
\end{figure}
\newpage
\twocolumngrid

This re-rise is governed by the interfacial conductance at the Si/SiO$_2$ interface, $h_{\mathrm{Si/SiO_2}}$, which we analyze using a modified two-temperature model (see Supplemental Materials for model details). To isolate the re-rise, we subtract the off-resonance data from the on-resonance data, as the feature arises uniquely from the SiO$_2$ response, which becomes significant when thermoreflectance is large at the optical phonon resonance. The corresponding fit and model sensitivity to $h_{\mathrm{Si/SiO_2}}$ are depicted in Fig. \ref{fig:1_PTR}c. The residual analysis indicates that the measurement constrains a lower bound of $\sim$160 MW\,m$^{-2}$\,K$^{-1}$ for $h_{\mathrm{Si/SiO_2}}$, while the limited change in residual at higher values precludes establishing a definitive upper bound.  This sensitivity to $h_{\mathrm{Si/SiO_2}}$ at optical phonon wavelengths demonstrates that infrared dielectric thermoreflectance provides insight into thermal transport phenomena that may not be resolved in conventional thermoreflectance configurations.  We note that prior reports of measured TBC across SiO$_2$/Si interfaces vary by over an order of magnitude, with the lower bounds ranging from $\sim$15-600 MW\,m$^{-2}$\,K$^{-1}$, complicating direct comparison of our results to prior works~\cite{zhuUltrafastThermoreflectanceTechniques2010,hopkinsCriteriaCrossPlaneDominated2010,kimlingThermalConductanceInterfaces2017,hurleyMeasurementKapitzaResistance2011}. 


This resonance-driven sensitivity also enables wavelength-selective optimization. By aligning the probe wavelength with TO phonon resonances, one can maximize the thermoreflectance coefficient, thereby improving the signal-to-noise ratio and enhancing sensitivity in thermal characterization techniques. At these peak wavelengths, $dR/dT$ for these dielectrics in the mid-infrared range surpasses that for commonly used metallic transducers~\cite{wilsonThermoreflectanceMetalTransducers2012b}. Specifically, as shown in Fig. \ref{fig:4_PTR}a, the thermoreflectance coefficients of sapphire, quartz, and AlN exceed the highest values reported for metallic transducers by about an order of magnitude, reaching peak $\lvert dR/dT\rvert$ values of $2.6\times10^{-3}$, $1.7\times10^{-3}$, and $2.1\times10^{-3}$ K$^{-1}$ near their corresponding TO phonon wavelengths of $\sim$21, $\sim$26, and $\sim$11 $\mu$m, respectively. In comparison, the thermoreflectance coefficient values for metal transducers reported in Ref.~\cite{wilsonThermoreflectanceMetalTransducers2012b} do not exceed $2.5\times10^{-4}$ K$^{-1}$. Unlike metals, which rely on broad interband transitions, dielectrics offer narrow, resonance-driven thermoreflectance peaks, enabling selective and highly sensitive thermal sensing.

While thermoreflectance peaks near TO phonon resonances, transducer performance depends not only on the thermoreflectance coefficient $dR/dT$ at the probe wavelength, but also on sufficient optical absorption at the pump wavelength to generate a measurable temperature modulation under fixed, low-fluence excitation. To provide a compact comparative metric, we define a figure of merit (FOM) as
\begin{equation}\label{Eq:FOM}
    FOM  = \alpha(\lambda_{pump})* \left| \frac{dR}{dT}(\lambda_{probe})\right|
\end{equation}

\noindent where $\alpha$ is the absorptivity of a thin film transducer. A dielectric thickness of 80 nm is chosen to represent a typical thermoreflectance transducer thickness \cite{braunSteadystateThermoreflectanceMethod2019,kohBulklikeIntrinsicPhonon2020,aryanaSuppressedElectronicContribution2021a,hopkinsCriteriaCrossPlaneDominated2010} and to enable direct comparison with literature-reported transducer FOMs. In these optically thin transducer configurations, heat deposition is volumetric and performance is governed by thermoreflectance sensitivity rather than heat-spreading limitations. In low-fluence thermoreflectance measurements, the detected signal scales primarily with $dR/dT$ at the probe wavelength, while absorptivity determines the pump fluence required to access that sensitivity.
This formulation ensures that maximizing the FOM leads to both large temperature rises within the transducer and strong thermoreflectance signals, optimizing the overall sensitivity of the measurement. This FOM serves as a compact screening metric for comparing transducer materials and identifying wavelength combinations that balance strong thermoreflectance sensitivity with sufficient pump absorption. This provides a consistent framework for comparing transducer performance across materials and wavelength combinations.

\onecolumngrid

\begin{figure}[b]
\centering
 \includegraphics[width=\textwidth]{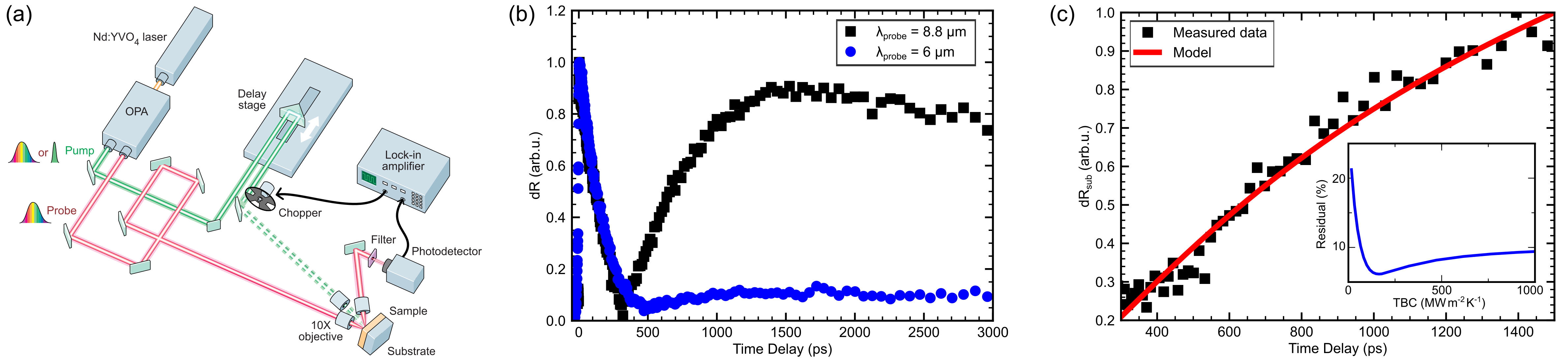}
    \caption[] {(a) Schematic of the transient thermoreflectance pump–probe setup. (b) Thermoreflectance response of a $\sim$100 nm thermally grown SiO$_2$ film on silicon measured with infrared probes at 8.8 $\mu$m (on resonance) and 6 $\mu$m (off resonance). (c) Modified two-temperature model fits to the differential signal ($dR_{sub} = dR_{8.8\mu\mathrm{ m}} - dR_{6\mu\mathrm{m}}$). The inset shows sensitivity to the Si/SiO$_2$ thermal boundary conductance, yielding a lower bound of $\sim160$~MW\,m$^{-2}$\,K$^{-1}$.}
  \label{fig:1_PTR}
\end{figure}
\newpage

\onecolumngrid

\begin{figure}
\centering
 \includegraphics[width=\textwidth]{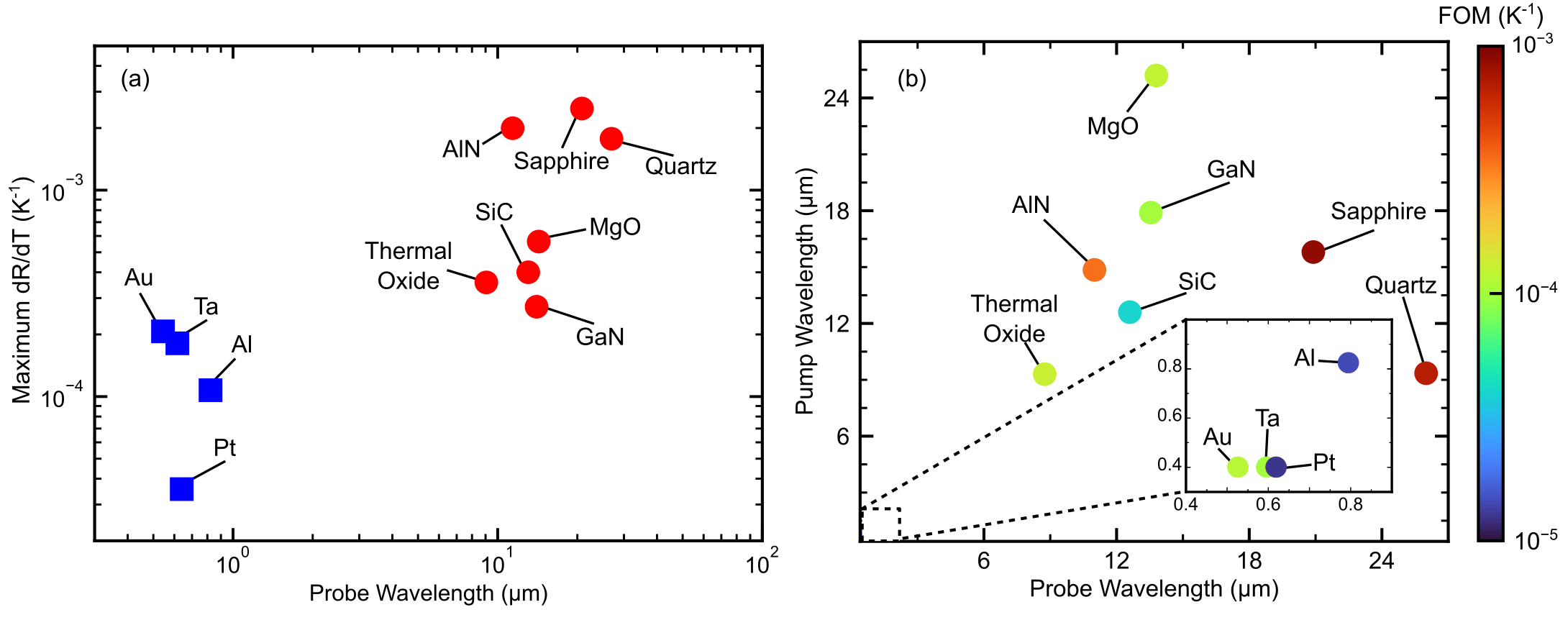}
    \caption[] {Computed maxima of (a) the thermoreflectance coefficient and (b) the FOM. Polar dielectric maxima from this work are compared with metal transducers from Ref.~\cite{wilsonThermoreflectanceMetalTransducers2012b}. The color axis in panel (b) indicates the maximum FOM magnitude for an 80-nm film of each material, evaluated at the corresponding pump and probe wavelengths. Optical properties of the metal transducers for absorptivity calculations were taken from Ref. \cite{palikHandbookOpticalConstants1998}. }
  \label{fig:4_PTR}
\end{figure}
\twocolumngrid

Following this approach, we can find the maximum FOMs for each prospective transducer studied here and in Ref.~\cite{wilsonThermoreflectanceMetalTransducers2012b}, as shown in Fig.~\ref{fig:4_PTR}b. The FOMs are plotted alongside their corresponding pump and probe wavelengths at which these maxima occur. Full FOM contour maps are available in the Supplemental Materials. As shown in Fig.~\ref{fig:4_PTR}b, the polar dielectrics analyzed in this study exhibit higher FOM values than common metallic transducers. Specifically, sapphire achieves a peak FOM of $8.7\times10^{-4}$, compared to $1.1\times10^{-4}$ for Au at 80 nm thickness. The spectral distribution of the FOM reflects distinct mechanisms in metallic and polar dielectric transducers. In metals, the FOM peaks in the visible to near-infrared, where interband absorption coincides with measurable thermoreflectance, then decreases at longer wavelengths as absorption weakens. In contrast, polar dielectrics peak in the mid- to far-infrared, where optical phonon resonances enhance both pump absorption and thermoreflectance via strong temperature-dependent optical constants. This separation highlights different transduction pathways and enables wavelength-selective thermoreflectance measurements, with pump–probe wavelengths chosen to exploit material-specific infrared resonances.

To summarize, we investigate the potential of dielectric materials as promising alternatives to traditional metallic transducers for thermoreflectance-based temperature sensing, particularly in the mid-infrared. Using infrared spectroscopic ellipsometry, we demonstrate that leveraging optical phonon resonances enables dielectric transducers to achieve thermoreflectance coefficients up to an order of magnitude higher than those of widely used metals, showcasing their ability to detect thermal variations.  Transient thermoreflectance measurements demonstrate that infrared dielectric thermoreflectance enables wavelength-selective access to thermal transport characteristics not readily accessible with conventional thermoreflectance. To compare materials across spectral regions, we introduce a figure of merit combining thermoreflectance response with pump absorption. Within this framework, dielectric materials such as sapphire and AlN exhibit up to an eightfold enhancement relative to conventional metal transducers, with the highest recorded value being $\sim$$10^{-3}$ K$^{-1}$ for sapphire. Moreover, the capability to fine-tune probe wavelengths to align with phonon dispersion unlocks new possibilities for refined thermal metrology, facilitating more accurate and targeted measurements in semiconductor diagnostics and in-situ material characterization. \newline

\centering {\textbf{Acknowledgments}}

\justifying
Research primarily supported as part of APEX (A Center for Power Electronics Materials and Manufacturing Exploration), an Energy Frontier Research Center funded by the U.S. Department of Energy (DOE), Office of Science, Basic Energy Sciences (BES), under Award \#ERW0345 and by the Office of Naval Research, Grant Number N00014-23-1-2630. 

\onecolumngrid
\vspace{2 cm}
\begin{spacing}{0.001}
\centering{\textbf{References}}
\end{spacing}
\twocolumngrid

\bibliography{PTR_main}

@article{medvedevElectronPhononCouplingSemiconductors2023,
  title = {Electron-Phonon Coupling in Semiconductors at High Electronic Temperatures},
  author = {Medvedev, Nikita},
  year = {2023},
  month = oct,
  journal = {Physical Review B},
  volume = {108},
  number = {14},
  pages = {144305},
  publisher = {American Physical Society},
  issn = {2469-9950},
  doi = {10.1103/PhysRevB.108.144305},
  urldate = {2026-02-19},
  copyright = {{\copyright} 2023 American Physical Society},
  langid = {english}
}

@article{aryanaSuppressedElectronicContribution2021a,
  title = {Suppressed Electronic Contribution in Thermal Conductivity of {{Ge2Sb2Se4Te}}},
  author = {Aryana, Kiumars and Zhang, Yifei and Tomko, John A. and Hoque, Md Shafkat Bin and Hoglund, Eric R. and Olson, David H. and Nag, Joyeeta and Read, John C. and R{\'i}os, Carlos and Hu, Juejun and Hopkins, Patrick E.},
  year = {2021},
  month = dec,
  journal = {Nature Communications},
  volume = {12},
  number = {1},
  pages = {7187},
  publisher = {Nature Publishing Group},
  issn = {2041-1723},
  doi = {10.1038/s41467-021-27121-x},
  urldate = {2025-02-23},
  copyright = {2021 The Author(s)},
  langid = {english}
}

@article{braunSteadystateThermoreflectanceMethod2019,
  title = {A Steady-State Thermoreflectance Method to Measure Thermal           Conductivity},
  author = {Braun, Jeffrey L. and Olson, David H. and Gaskins, John T. and Hopkins, Patrick E.},
  year = {2019},
  month = feb,
  journal = {Review of Scientific Instruments},
  volume = {90},
  number = {2},
  pages = {024905},
  publisher = {AIP Publishing LLC AIP Publishing},
  issn = {0034-6748},
  doi = {10.1063/1.5056182},
  urldate = {2023-02-25},
  pmid = {30831683}
}

@article{bridaGenerationBroadbandMidinfrared2007,
  title = {Generation of Broadband Mid-Infrared Pulses from an Optical Parametric Amplifier},
  author = {Brida, D. and Manzoni, C. and Cirmi, G. and Marangoni, M. and Silvestri, S. De and Cerullo, G.},
  year = {2007},
  month = nov,
  journal = {Optics Express},
  volume = {15},
  number = {23},
  pages = {15035--15040},
  publisher = {Optica Publishing Group},
  issn = {1094-4087},
  doi = {10.1364/OE.15.015035},
  urldate = {2025-03-08},
  copyright = {{\copyright} 2007 Optical Society of America},
  langid = {english}
}

@article{cahillAnalysisHeatFlow2004,
  title = {Analysis of Heat Flow in Layered Structures for Time-Domain Thermoreflectance},
  author = {Cahill, David G.},
  year = {2004},
  month = dec,
  journal = {Review of Scientific Instruments},
  volume = {75},
  number = {12},
  pages = {5119--5122},
  issn = {00346748},
  doi = {10.1063/1.1819431}
}

@article{choiIndirectHeatingPt2014,
  title = {Indirect Heating of {{Pt}} by Short-Pulse Laser Irradiation of {{Au}} in a Nanoscale {{Pt}}/{{Au}} Bilayer},
  author = {Choi, Gyung-Min and Wilson, R. B. and Cahill, David G.},
  year = {2014},
  month = feb,
  journal = {Physical Review B},
  volume = {89},
  number = {6},
  pages = {064307},
  publisher = {American Physical Society},
  doi = {10.1103/PhysRevB.89.064307},
  urldate = {2024-09-17}
}

@article{choiThermalSpintransferTorque2015b,
  title = {Thermal Spin-Transfer Torque Driven by the Spin-Dependent {{Seebeck}} Effect in Metallic Spin-Valves},
  author = {Choi, Gyung-Min and Moon, Chul-Hyun and Min, Byoung-Chul and Lee, Kyung-Jin and Cahill, David G.},
  year = {2015},
  month = jul,
  journal = {Nature Physics},
  volume = {11},
  number = {7},
  pages = {576--581},
  publisher = {Nature Publishing Group},
  issn = {1745-2481},
  doi = {10.1038/nphys3355},
  urldate = {2025-05-08},
  copyright = {2014 Springer Nature Limited},
  langid = {english}
}

@article{colavitaThermoreflectanceTestMo1983,
  title = {Thermoreflectance Test of {{W}}, {{Mo}}, and Paramagnetic {{Cr}} Band Structures},
  author = {Colavita, E. and Franciosi, A. and Mariani, C. and Rosei, R.},
  year = {1983},
  month = apr,
  journal = {Physical Review B},
  volume = {27},
  number = {8},
  pages = {4684--4693},
  publisher = {American Physical Society},
  doi = {10.1103/PhysRevB.27.4684},
  urldate = {2025-02-23}
}

@article{delfyettHighpowerUltrafastLaser1992,
  title = {High-Power Ultrafast Laser Diodes},
  author = {Delfyett, P.J. and Florez, L.T. and Stoffel, N. and Gmitter, T. and Andreadakis, N.C. and Silberberg, Y. and Heritage, J.P. and Alphonse, G.A.},
  year = {1992},
  month = oct,
  journal = {IEEE Journal of Quantum Electronics},
  volume = {28},
  number = {10},
  pages = {2203--2219},
  issn = {1558-1713},
  doi = {10.1109/3.159528},
  urldate = {2025-03-08}
}

@article{dumkeInterbandTransitionsMaser1962,
  title = {Interband {{Transitions}} and {{Maser Action}}},
  author = {Dumke, W. P.},
  year = {1962},
  month = sep,
  journal = {Physical Review},
  volume = {127},
  number = {5},
  pages = {1559--1563},
  issn = {0031-899X},
  doi = {10.1103/PhysRev.127.1559},
  urldate = {2025-02-24},
  copyright = {http://link.aps.org/licenses/aps-default-license},
  langid = {english}
}

@article{feserProbingAnisotropicHeat2012,
  title = {Probing Anisotropic Heat Transport Using Time-Domain Thermoreflectance with Offset Laser Spots},
  author = {Feser, Joseph P. and Cahill, David G.},
  year = {2012},
  month = oct,
  journal = {Review of Scientific Instruments},
  volume = {83},
  number = {10},
  pages = {104901},
  issn = {0034-6748, 1089-7623},
  doi = {10.1063/1.4757863},
  urldate = {2024-06-13},
  langid = {english}
}

@article{gambettaRealtimeObservationNonlinear2006,
  title = {Real-Time Observation of Nonlinear Coherent Phonon Dynamics in Single-Walled Carbon Nanotubes},
  author = {Gambetta, A. and Manzoni, C. and Menna, E. and Meneghetti, M. and Cerullo, G. and Lanzani, G. and Tretiak, S. and Piryatinski, A. and Saxena, A. and Martin, R. L. and Bishop, A. R.},
  year = {2006},
  month = aug,
  journal = {Nature Physics},
  volume = {2},
  number = {8},
  pages = {515--520},
  publisher = {Nature Publishing Group},
  issn = {1745-2481},
  doi = {10.1038/nphys345},
  urldate = {2025-05-08},
  copyright = {2006 Springer Nature Limited},
  langid = {english}
}

@article{giriMechanismsNonequilibriumElectronphonon2015a,
  title = {Mechanisms of Nonequilibrium Electron-Phonon Coupling and Thermal Conductance at Interfaces},
  author = {Giri, Ashutosh and Gaskins, John T. and Donovan, Brian F. and Szwejkowski, Chester and Warzoha, Ronald J. and Rodriguez, Mark A. and Ihlefeld, Jon and Hopkins, Patrick E.},
  year = {2015},
  month = mar,
  journal = {Journal of Applied Physics},
  volume = {117},
  number = {10},
  pages = {105105},
  issn = {0021-8979},
  doi = {10.1063/1.4914867},
  urldate = {2024-10-07}
}

@article{grafNodalQuasiparticleMeltdown2011,
  title = {Nodal Quasiparticle Meltdown in Ultrahigh-Resolution Pump--Probe Angle-Resolved Photoemission},
  author = {Graf, J. and Jozwiak, C. and Smallwood, C. L. and Eisaki, H. and Kaindl, R. A. and Lee, D.-H. and Lanzara, A.},
  year = {2011},
  month = oct,
  journal = {Nature Physics},
  volume = {7},
  number = {10},
  pages = {805--809},
  publisher = {Nature Publishing Group},
  issn = {1745-2481},
  doi = {10.1038/nphys2027},
  urldate = {2025-05-08},
  copyright = {2011 Springer Nature Limited},
  langid = {english}
}

@article{halteOpticalResponsePeriodically2006,
  title = {Optical Response of Periodically Modulated Nanostructures near the Interband Transition Threshold of Noble Metals},
  author = {Halt{\'e}, V. and Benabbas, A. and Bigot, J.-Y.},
  year = {2006},
  month = apr,
  journal = {Optics Express},
  volume = {14},
  number = {7},
  pages = {2909--2920},
  publisher = {Optica Publishing Group},
  issn = {1094-4087},
  doi = {10.1364/OE.14.002909},
  urldate = {2025-02-24},
  copyright = {{\copyright} 2006 Optical Society of America},
  langid = {english}
}

@article{hirtIncreasedThermalConductivity2024b,
  title = {Increased Thermal Conductivity and Decreased Electron--Phonon Coupling Factor of the Aluminum Scandium Intermetallic Phase ({{Al3Sc}}) Compared to Solid Solutions},
  author = {Hirt, Daniel and Islam, Md. Rafiqul and Hoque, Md. Shafkat Bin and Hutchins, William and Makarem, Sara and Lenox, Megan K. and Riffe, William T. and Ihlefeld, Jon F. and Scott, Ethan A. and Esteves, Giovanni and Hopkins, Patrick E.},
  year = {2024},
  month = may,
  journal = {Applied Physics Letters},
  volume = {124},
  number = {20},
  pages = {202202},
  issn = {0003-6951},
  doi = {10.1063/5.0201763},
  urldate = {2025-02-23}
}

@article{hopkinsCriteriaCrossPlaneDominated2010,
  title = {Criteria for {{Cross-Plane Dominated Thermal Transport}} in {{Multilayer Thin Film Systems During Modulated Laser Heating}}},
  author = {Hopkins, Patrick E. and Serrano, Justin R. and Phinney, Leslie M. and Kearney, Sean P. and Grasser, Thomas W. and Harris, C. Thomas},
  year = {2010},
  month = aug,
  journal = {Journal of Heat Transfer},
  volume = {132},
  number = {8},
  pages = {081302},
  issn = {0022-1481, 1528-8943},
  doi = {10.1115/1.4000993},
  urldate = {2026-02-11},
  langid = {english}
}

@article{hoqueExperimentalObservationBallistic2024,
  title = {Experimental Observation of Ballistic to Diffusive Transition in Phonon Thermal Transport of {{AlN}} Thin Films},
  author = {Hoque, Md Shafkat Bin and Koh, Yee Rui and Zare, Saman and Liao, Michael E. and Huynh, Kenny and Goorsky, Mark S. and Liu, Zeyu and Shi, Jingjing and Graham, Samuel and Luo, Tengfei and Ahmad, Habib and Doolittle, W. Alan and Hopkins, Patrick E.},
  year = {2024},
  month = dec,
  journal = {Applied Physics Letters},
  volume = {125},
  number = {26},
  pages = {262201},
  issn = {0003-6951},
  doi = {10.1063/5.0239769},
  urldate = {2025-02-23}
}

@article{hurleyMeasurementKapitzaResistance2011,
  title = {Measurement of the {{Kapitza}} Resistance across a Bicrystal Interface},
  author = {Hurley, D. H. and Khafizov, M. and Shinde, S. L.},
  year = {2011},
  month = apr,
  journal = {Journal of Applied Physics},
  volume = {109},
  number = {8},
  pages = {083504},
  issn = {0021-8979},
  doi = {10.1063/1.3573511},
  urldate = {2026-02-11}
}

@article{hutchinsUltrafastEvanescentHeat2025,
  title = {Ultrafast Evanescent Heat Transfer across Solid Interfaces via Hyperbolic Phonon--Polariton Modes in Hexagonal Boron Nitride},
  author = {Hutchins, William and Zare, Saman and Hirt, Dan M. and Tomko, John A. and Matson, Joseph R. and {Diaz-Granados}, Katja and Long, Mackey and He, Mingze and Pfeifer, Thomas and Li, Jiahan and Edgar, James H. and Maria, Jon-Paul and Caldwell, Joshua D. and Hopkins, Patrick E.},
  year = {2025},
  month = mar,
  journal = {Nature Materials},
  pages = {1--9},
  publisher = {Nature Publishing Group},
  issn = {1476-4660},
  doi = {10.1038/s41563-025-02154-5},
  urldate = {2025-03-23},
  copyright = {2025 The Author(s)},
  langid = {english}
}

@article{jamgotchianSituObservationInterferometric2001,
  title = {In Situ Observation and Interferometric Characterization of Solid--Liquid Interface Morphology in Directionally Growing Transparent Model Systems},
  author = {Jamgotchian, H. and Bergeon, N. and Benielli, D. and Voge, P. and Billia, B.},
  year = {2001},
  journal = {Journal of Microscopy},
  volume = {203},
  number = {1},
  pages = {119--127},
  issn = {1365-2818},
  doi = {10.1046/j.1365-2818.2001.00900.x},
  urldate = {2025-02-24},
  langid = {english}
}

@article{kennesTransientSuperconductivityElectronic2017,
  title = {Transient Superconductivity from Electronic Squeezing of Optically Pumped Phonons},
  author = {Kennes, Dante M. and Wilner, Eli Y. and Reichman, David R. and Millis, Andrew J.},
  year = {2017},
  month = may,
  journal = {Nature Physics},
  volume = {13},
  number = {5},
  pages = {479--483},
  publisher = {Nature Publishing Group},
  issn = {1745-2481},
  doi = {10.1038/nphys4024},
  urldate = {2025-05-08},
  copyright = {2017 Springer Nature Limited},
  langid = {english}
}

@article{kimlingThermalConductanceInterfaces2017,
  title = {Thermal Conductance of Interfaces with Amorphous {{SiO}} 2 Measured by Time-Resolved Magneto-Optic {{Kerr-effect}} Thermometry},
  author = {Kimling, Judith and {Philippi-Kobs}, Andr{\'e} and Jacobsohn, Jonathan and Oepen, Hans Peter and Cahill, David G.},
  year = {2017},
  month = may,
  journal = {Physical Review B},
  volume = {95},
  number = {18},
  pages = {184305},
  issn = {2469-9950, 2469-9969},
  doi = {10.1103/PhysRevB.95.184305},
  urldate = {2026-02-11},
  copyright = {http://link.aps.org/licenses/aps-default-license},
  langid = {english}
}

@article{kohBulklikeIntrinsicPhonon2020,
  title = {Bulk-like {{Intrinsic Phonon Thermal Conductivity}} of {{Micrometer-Thick AlN Films}}},
  author = {Koh, Yee Rui and Cheng, Zhe and Mamun, Abdullah and Bin Hoque, Md Shafkat and Liu, Zeyu and Bai, Tingyu and Hussain, Kamal and Liao, Michael E. and Li, Ruiyang and Gaskins, John T. and Giri, Ashutosh and Tomko, John and Braun, Jeffrey L. and Gaevski, Mikhail and Lee, Eungkyu and Yates, Luke and Goorsky, Mark S. and Luo, Tengfei and Khan, Asif and Graham, Samuel and Hopkins, Patrick E.},
  year = {2020},
  month = jul,
  journal = {ACS Applied Materials \& Interfaces},
  volume = {12},
  number = {26},
  pages = {29443--29450},
  publisher = {American Chemical Society},
  issn = {1944-8244},
  doi = {10.1021/acsami.0c03978},
  urldate = {2025-03-27}
}

@article{kohHighThermalConductivity2021,
  title = {High Thermal Conductivity and Thermal Boundary Conductance of Homoepitaxially Grown Gallium Nitride ({{GaN}}) Thin Films},
  author = {Koh, Yee Rui and Hoque, Md Shafkat Bin and Ahmad, Habib and Olson, David H. and Liu, Zeyu and Shi, Jingjing and Wang, Yekan and Huynh, Kenny and Hoglund, Eric R. and Aryana, Kiumars and Howe, James M. and Goorsky, Mark S. and Graham, Samuel and Luo, Tengfei and Hite, Jennifer K. and Doolittle, W. Alan and Hopkins, Patrick E.},
  year = {2021},
  month = oct,
  journal = {Physical Review Materials},
  volume = {5},
  number = {10},
  pages = {104604},
  publisher = {American Physical Society},
  doi = {10.1103/PhysRevMaterials.5.104604},
  urldate = {2025-03-27}
}

@article{kuzmenkoUniversalOpticalConductance2008,
  title = {Universal {{Optical Conductance}} of {{Graphite}}},
  author = {Kuzmenko, A. B. and {van Heumen}, E. and Carbone, F. and {van der Marel}, D.},
  year = {2008},
  month = mar,
  journal = {Physical Review Letters},
  volume = {100},
  number = {11},
  pages = {117401},
  publisher = {American Physical Society},
  doi = {10.1103/PhysRevLett.100.117401},
  urldate = {2025-02-24}
}

@book{lorentzTheoryElectronsIts1909,
  title = {The {{Theory}} of {{Electrons}} and {{Its Applications}} to the {{Phenomena}} of {{Light}} and {{Radiant Heat}}: {{A Course}} of {{Lectures Delivered}} in {{Columbia University}}, {{New York}}, in {{March}} and {{April}}, 1906},
  shorttitle = {The {{Theory}} of {{Electrons}} and {{Its Applications}} to the {{Phenomena}} of {{Light}} and {{Radiant Heat}}},
  author = {Lorentz, Hendrik Antoon},
  year = {1909},
  publisher = {B.G. Teubner},
  googlebooks = {DGFDAAAAIAAJ},
  langid = {english}
}

@article{morgnerSubtwocyclePulsesKerrlens1999,
  title = {Sub-Two-Cycle Pulses from a {{Kerr-lens}} Mode-Locked {{Ti}}:Sapphire Laser},
  shorttitle = {Sub-Two-Cycle Pulses from a {{Kerr-lens}} Mode-Locked {{Ti}}},
  author = {Morgner, U. and K{\"a}rtner, F. X. and Cho, S. H. and Chen, Y. and Haus, H. A. and Fujimoto, J. G. and Ippen, E. P. and Scheuer, V. and Angelow, G. and Tschudi, T.},
  year = {1999},
  month = mar,
  journal = {Optics Letters},
  volume = {24},
  number = {6},
  pages = {411--413},
  publisher = {Optica Publishing Group},
  issn = {1539-4794},
  doi = {10.1364/OL.24.000411},
  urldate = {2025-03-08},
  copyright = {{\copyright} 1999 Optical Society of America},
  langid = {english}
}

@article{otterTemperaturabhaengigkeitOptischenKonstanten1961,
  title = {{Temperaturabh{\"a}ngigkeit der optischen Konstanten massiver Metalle}},
  author = {Otter, M.},
  year = {1961},
  month = oct,
  journal = {Zeitschrift f{\"u}r Physik},
  volume = {161},
  number = {5},
  pages = {539--549},
  issn = {0044-3328},
  doi = {10.1007/BF01341551},
  urldate = {2025-02-23},
  langid = {ngerman}
}

@book{palikHandbookOpticalConstants1998,
  title = {Handbook of {{Optical Constants}} of {{Solids}}},
  author = {Palik, Edward D.},
  year = {1998},
  publisher = {Academic Press},
  googlebooks = {nxoqxyoHfbIC},
  isbn = {978-0-12-544423-1},
  langid = {english}
}

@article{roseiElectronicStructureBcc1980,
  title = {Electronic Structure of the Bcc Transition Metals: {{Thermoreflectance}} Studies of Bulk {{V}}, {{Nb}}, {{Ta}}, and Alpha {{TaHx}}\$},
  shorttitle = {Electronic Structure of the Bcc Transition Metals},
  author = {Rosei, R. and Colavita, E. and Franciosi, A. and Weaver, J. H. and Peterson, D. T.},
  year = {1980},
  month = apr,
  journal = {Physical Review B},
  volume = {21},
  number = {8},
  pages = {3152--3157},
  publisher = {American Physical Society},
  doi = {10.1103/PhysRevB.21.3152},
  urldate = {2025-02-23}
}

@article{roseiThermomodulationSpectraAu1972,
  title = {Thermomodulation {{Spectra}} of {{Al}}, {{Au}}, and {{Cu}}},
  author = {Rosei, R. and Lynch, D. W.},
  year = {1972},
  month = may,
  journal = {Physical Review B},
  volume = {5},
  number = {10},
  pages = {3883--3894},
  publisher = {American Physical Society},
  doi = {10.1103/PhysRevB.5.3883},
  urldate = {2025-02-23}
}

@article{schmidtCharacterizationThinMetal2010,
  title = {Characterization of Thin Metal Films via Frequency-Domain Thermoreflectance},
  author = {Schmidt, Aaron J. and Cheaito, Ramez and Chiesa, Matteo},
  year = {2010},
  month = jan,
  journal = {Journal of Applied Physics},
  volume = {107},
  number = {2},
  pages = {024908},
  issn = {0021-8979, 1089-7550},
  doi = {10.1063/1.3289907},
  urldate = {2024-06-13},
  langid = {english}
}

@article{scoulerTemperatureModulatedReflectanceGold1967,
  title = {Temperature-{{Modulated Reflectance}} of {{Gold}} from 2 to 10 {{eV}}},
  author = {Scouler, W. J.},
  year = {1967},
  month = mar,
  journal = {Physical Review Letters},
  volume = {18},
  number = {12},
  pages = {445--448},
  publisher = {American Physical Society},
  doi = {10.1103/PhysRevLett.18.445},
  urldate = {2025-02-23}
}

@article{tomkoLonglivedModulationPlasmonic2021,
  title = {Long-Lived Modulation of Plasmonic Absorption by Ballistic Thermal Injection},
  author = {Tomko, John A. and Runnerstrom, Evan L. and Wang, Yi-Siang and Chu, Weibin and Nolen, Joshua R. and Olson, David H. and Kelley, Kyle P. and Cleri, Angela and Nordlander, Josh and Caldwell, Joshua D. and Prezhdo, Oleg V. and Maria, Jon-Paul and Hopkins, Patrick E.},
  year = {2021},
  month = jan,
  journal = {Nature Nanotechnology},
  volume = {16},
  number = {1},
  pages = {47--51},
  publisher = {Nature Publishing Group},
  issn = {1748-3395},
  doi = {10.1038/s41565-020-00794-z},
  urldate = {2025-03-24},
  copyright = {2020 The Author(s), under exclusive licence to Springer Nature Limited},
  langid = {english}
}

@article{tomkoUltrafastChargeCarrier2025,
  title = {Ultrafast {{Charge Carrier Dynamics}} in {{Vanadium Dioxide}}, {{VO2}}: {{Nonequilibrium Contributions}} to the {{Photoinduced Phase Transitions}}},
  shorttitle = {Ultrafast {{Charge Carrier Dynamics}} in {{Vanadium Dioxide}}, {{VO2}}},
  author = {Tomko, John A. and Aryana, Kiumars and Wu, Yifan and Zhou, Guoqing and Zhang, Qiyan and Wongwiset, Pat and Wheeler, Virginia and Prezhdo, Oleg V. and Hopkins, Patrick E.},
  year = {2025},
  month = feb,
  journal = {The Journal of Physical Chemistry Letters},
  volume = {16},
  number = {5},
  pages = {1312--1319},
  publisher = {American Chemical Society},
  doi = {10.1021/acs.jpclett.4c02951},
  urldate = {2025-03-24}
}

@article{tzallasExtremeultravioletPumpProbe2011,
  title = {Extreme-Ultraviolet Pump--Probe Studies of One-Femtosecond-Scale Electron Dynamics},
  author = {Tzallas, P. and Skantzakis, E. and Nikolopoulos, L. a. A. and Tsakiris, G. D. and Charalambidis, D.},
  year = {2011},
  month = oct,
  journal = {Nature Physics},
  volume = {7},
  number = {10},
  pages = {781--784},
  publisher = {Nature Publishing Group},
  issn = {1745-2481},
  doi = {10.1038/nphys2033},
  urldate = {2025-05-08},
  copyright = {2011 Springer Nature Limited},
  langid = {english}
}

@book{willardsonModulationTechniques1972,
  title = {Modulation {{Techniques}}},
  author = {Willardson, Robert K. and Beer, Albert C.},
  year = {1972},
  publisher = {Academic Press},
  googlebooks = {sd8\_xR1ZuOwC},
  isbn = {978-0-12-752109-1},
  langid = {english}
}

@article{wilsonThermoreflectanceMetalTransducers2012b,
  title = {Thermoreflectance of Metal Transducers for Optical Pump-Probe Studies of Thermal Properties},
  author = {Wilson, R. B. and Apgar, Brent A. and Martin, Lane W. and Cahill, David G.},
  year = {2012},
  month = dec,
  journal = {Optics Express},
  volume = {20},
  number = {27},
  pages = {28829--28838},
  publisher = {Optica Publishing Group},
  issn = {1094-4087},
  doi = {10.1364/OE.20.028829},
  urldate = {2025-02-24},
  copyright = {{\copyright} 2012 OSA},
  langid = {english}
}

@article{zhangUltrafastRelaxationLattice2023,
  title = {Ultrafast Relaxation of Lattice Distortion in Two-Dimensional Perovskites},
  author = {Zhang, Hao and Li, Wenbin and Essman, Joseph and Quarti, Claudio and Metcalf, Isaac and Chiang, Wei-Yi and Sidhik, Siraj and Hou, Jin and Fehr, Austin and Attar, Andrew and Lin, Ming-Fu and Britz, Alexander and Shen, Xiaozhe and Link, Stephan and Wang, Xijie and Bergmann, Uwe and Kanatzidis, Mercouri G. and Katan, Claudine and Even, Jacky and Blancon, Jean-Christophe and Mohite, Aditya D.},
  year = {2023},
  month = apr,
  journal = {Nature Physics},
  volume = {19},
  number = {4},
  pages = {545--550},
  publisher = {Nature Publishing Group},
  issn = {1745-2481},
  doi = {10.1038/s41567-022-01903-6},
  urldate = {2025-05-08},
  copyright = {2023 The Author(s), under exclusive licence to Springer Nature Limited},
  langid = {english}
}

@article{zhuUltrafastThermoreflectanceTechniques2010,
  title = {Ultrafast Thermoreflectance Techniques for Measuring Thermal Conductivity and Interface Thermal Conductance of Thin Films},
  author = {Zhu, Jie and Tang, Dawei and Wang, Wei and Liu, Jun and Holub, Kristopher W. and Yang, Ronggui},
  year = {2010},
  month = nov,
  journal = {Journal of Applied Physics},
  volume = {108},
  number = {9},
  pages = {094315},
  issn = {0021-8979},
  doi = {10.1063/1.3504213},
  urldate = {2026-02-11}
}

\onecolumngrid

\newpage

\onecolumngrid

\title[SM]{Supplemental Materials: ``Phonon-Driven Thermoreflectance in Polar Dielectrics''}

\affiliation{$Department~of~Mechanical~and~Aerospace~Engineering,~University~of~Virginia,~Charlottesville,~Virginia~22904, USA$}
\affiliation{$Department~of~Materials~Science~and~Engineering,~University~of~Virginia,~Charlottesville,~Virginia~22904, USA$}
\affiliation{$Department~of~Physics,~University~of~Virginia,~Charlottesville,~Virginia~22904, USA$}

\author{William D. Hutchins}
\thanks{These two authors contributed equally.}
\author{Saman Zare}
\thanks{These two authors contributed equally.}
\author{Daniel Hirt}
\author{Elizabeth Golightly}

\affiliation{$Department~of~Mechanical~and~Aerospace~Engineering,~University~of~Virginia,~Charlottesville,~Virginia~22904, USA$}

\author{Patrick E. Hopkins}
\email{Corresponding Author: phopkins@virginia.edu}
\affiliation{$Department~of~Mechanical~and~Aerospace~Engineering,~University~of~Virginia,~Charlottesville,~Virginia~22904, USA$}
\affiliation{$Department~of~Materials~Science~and~Engineering,~University~of~Virginia,~Charlottesville,~Virginia~22904, USA$}
\affiliation{$Department~of~Physics,~University~of~Virginia,~Charlottesville,~Virginia~22904, USA$}

\maketitle

\setcounter{equation}{0}
\setcounter{figure}{0}
\setcounter{table}{0}
\setcounter{page}{1}
\makeatletter
\renewcommand{\theequation}{S\arabic{equation}}
\renewcommand{\thefigure}{S\arabic{figure}}
\renewcommand{\bibnumfmt}[1]{[#1]}
\renewcommand{\citenumfont}[1]{#1}

\onecolumngrid
{ \section*{Supplemental Materials} }
\maketitle
\section*{Section A: Sample Details}

The samples used in this study include both epitaxially grown thin films and commercially available bulk materials. Aluminum nitride (AlN) and gallium nitride (GaN) were each deposited on sapphire substrates with a thickness of approximately 2.5 $\mu$m. The AlN layer was grown using the metal modulated epitaxy (MME) method while the GaN layer was deposited via the metal-organic chemical vapor deposition (MOCVD) technique, as detailed in Refs. \cite{kohBulklikeIntrinsicPhonon2020} and \cite{kohHighThermalConductivity2021}, respectively. These high-quality epitaxial films were prepared to ensure consistency in material properties for accurate comparison. A $\sim$100 nm thermally grown SiO$_2$ film on silicon, obtained from University Wafer (sample number), was included and characterized by spectroscopic ellipsometry and transient thermoreflectance measurements. In addition, we used bulk samples (with a thickness of 1 mm) acquired from commercial vendors for magnesium oxide (MgO, MTI Corporation, MGa101005S1), 4H-silicon carbide (4H-SiC, MSE Supplies, WA0333), sapphire (MSE Supplies), and quartz (University Wafer, 3676). These substrates were selected to represent a diverse range of optical and structural properties relevant to the objectives of this study.

\section*{Section B: Spectroscopic Ellipsometry Measurements, Model Fitting, and Uncertainty Quantification}

Infrared spectroscopic ellipsometry was used to extract the wavelength-dependent complex refractive index of all dielectric samples studied in this work. The ellipsometric parameters $\Psi$ and $\Delta$ were measured as a function of wavelength using a J.A.~Woollam IR-VASE system over the spectral range relevant to the infrared-active optical phonon modes. Measurements were performed at two angles of incidence (50\textdegree and 60\textdegree) to improve sensitivity to the optical response and to reduce parameter correlation in the fitting procedure. For temperature-dependent measurements, samples were mounted on a temperature-controlled heating stage and allowed to thermally equilibrate prior to data acquisition. Ellipsometry measurements were conducted at two temperatures, 25~$^\circ$C and 200~$^\circ$C, for all samples.

To extract optical constants from the measured ellipsometric data, the $\Psi$ and $\Delta$ spectra were analyzed using optical models appropriate to each sample geometry. Bulk substrates were modeled as semi-infinite media, while thin-film samples were represented as a single dielectric layer on a sapphire substrate. In all cases, the dielectric response was described using a Lorentz oscillator formalism to capture the infrared-active optical phonon resonances that dominate the optical behavior in this spectral range. Model parameters were obtained through regression analysis by simultaneously fitting $\Psi$ and $\Delta$ across all measured angles of incidence, ensuring consistency across the dataset.

\begin{figure}
 \includegraphics[width=0.9\textwidth]{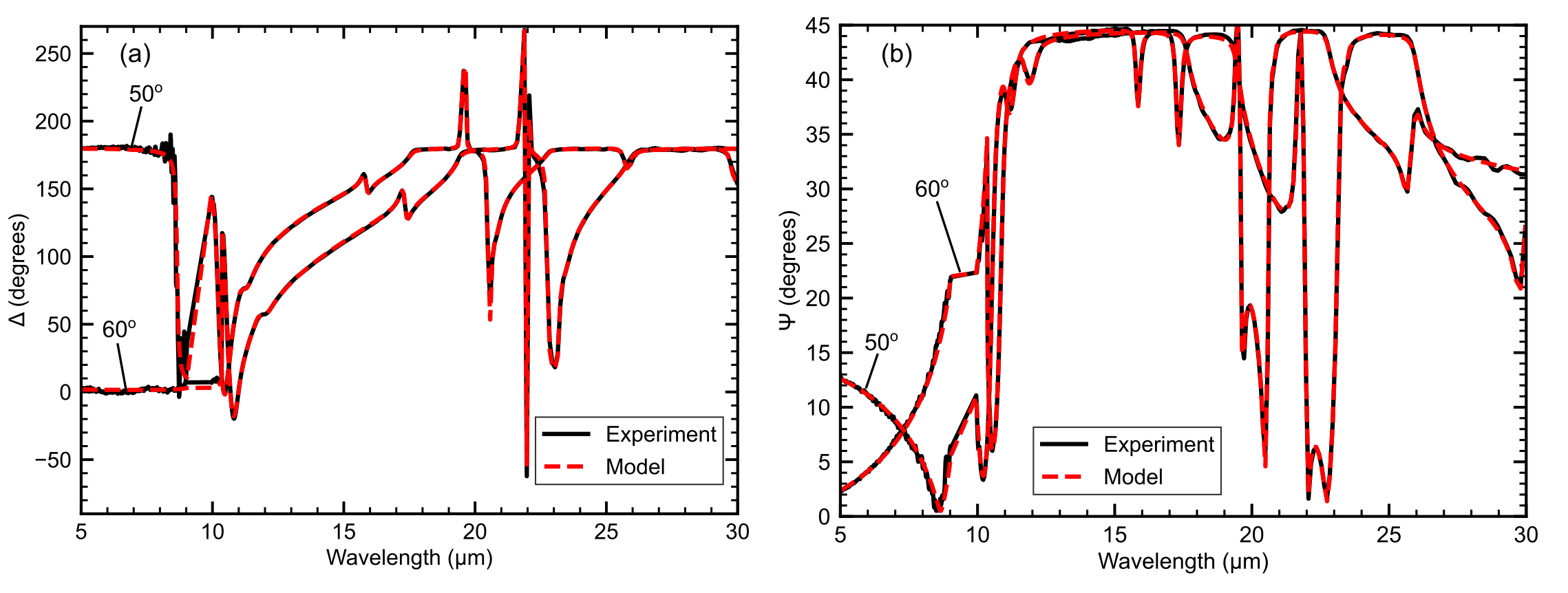}
    \caption[] {Representative ellipsometry data measured at incidence angles of 50\textdegree and 60\textdegree are shown together with the corresponding model fits to the measured data, with sapphire shown here as an example sample.}
  \label{fig:S1_PTR}
\end{figure}

The quality of the optical modeling and fitting procedure is illustrated in Fig.~\ref{fig:S1_PTR}, which shows representative ellipsometry data for sapphire, including the measured $\Psi$ and $\Delta$ spectra together with the corresponding model fits at multiple angles of incidence. The close agreement between experiment and model across the full spectral range demonstrates that the Lorentz-based dielectric function accurately captures the phonon-mediated optical response. This agreement provides confidence that the extracted complex refractive indices are physically meaningful and reliably determined.

\begin{figure}
 \includegraphics[width=0.9\textwidth]{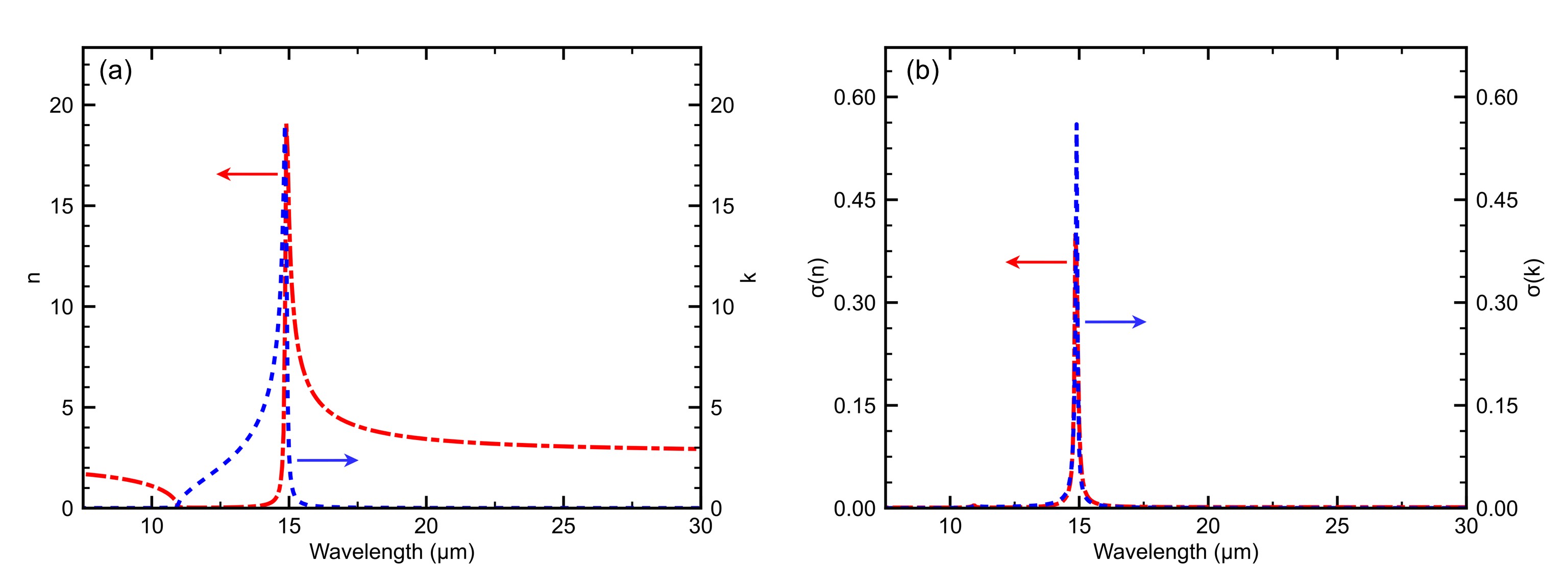}
    \caption[] {(a) Real (\textit{n}) and imaginary (\textit{k}) parts of the in-plane complex refractive index of AlN at room temperature, extracted from spectroscopic ellipsometry measurements. (b) Corresponding standard deviations of \textit{n} and \textit{k}, representing the measurement uncertainty associated with the ellipsometric fitting.}
  \label{fig:S2_PTR}
\end{figure}

\begin{figure}
 \includegraphics[width=0.9\textwidth]{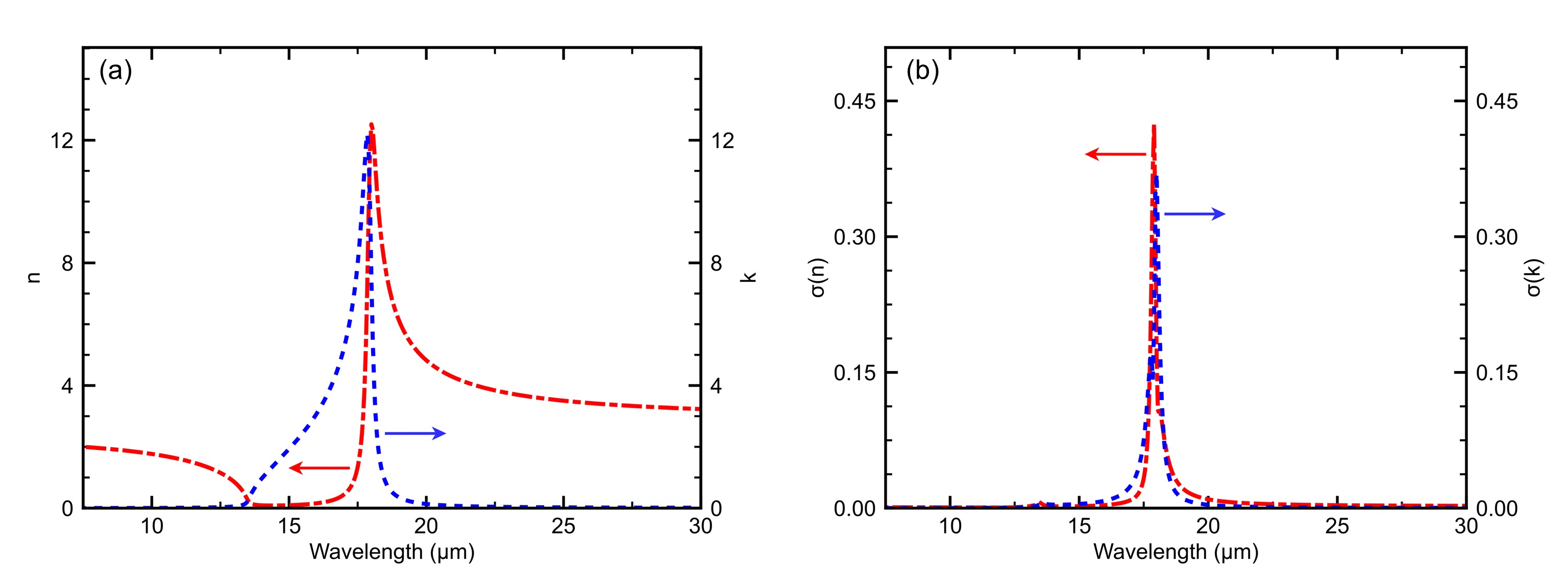}
    \caption[] {(a) Real (\textit{n}) and imaginary (\textit{k}) parts of the in-plane complex refractive index of GaN at room temperature, extracted from spectroscopic ellipsometry measurements. (b) Corresponding standard deviations of \textit{n} and \textit{k}, representing the measurement uncertainty associated with the ellipsometric fitting.}
  \label{fig:S3_PTR}
\end{figure}

\begin{figure}
 \includegraphics[width=0.9\textwidth]{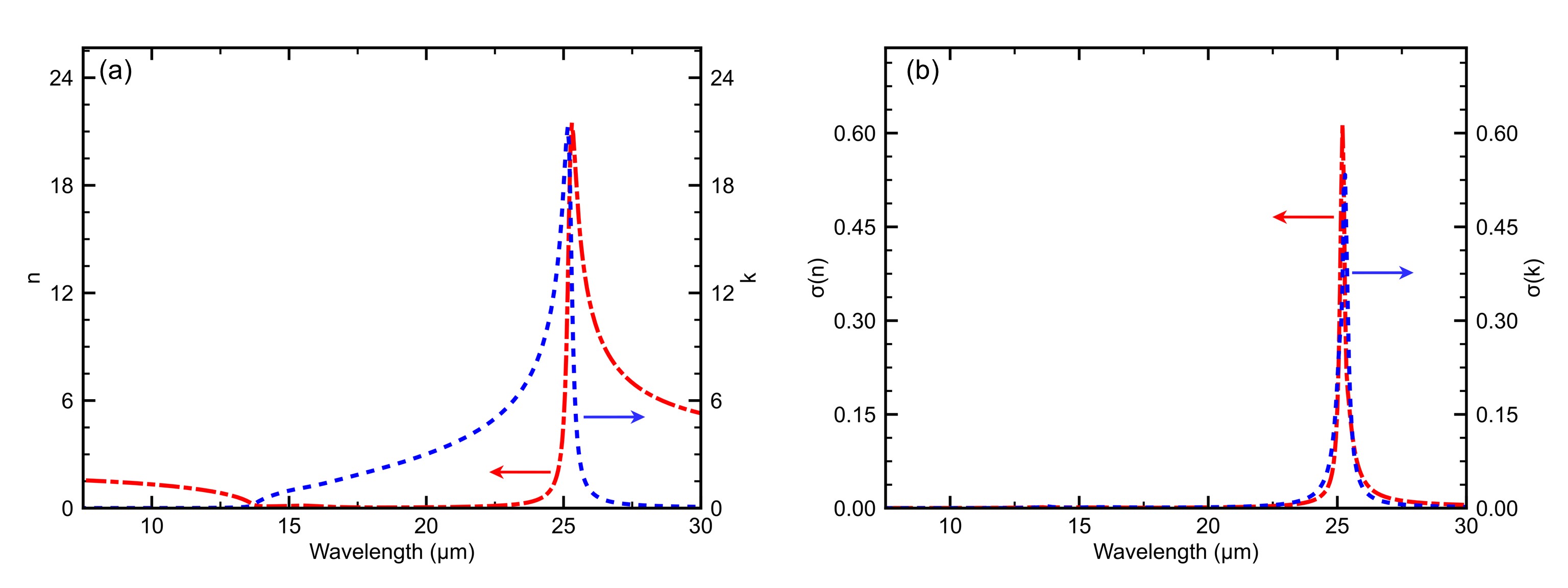}
    \caption[] {(a) Real (\textit{n}) and imaginary (\textit{k}) parts of the complex refractive index of MgO at room temperature, extracted from spectroscopic ellipsometry measurements. (b) Corresponding standard deviations of \textit{n} and \textit{k}, representing the measurement uncertainty associated with the ellipsometric fitting.}
  \label{fig:S4_PTR}
\end{figure}

\begin{figure}
 \includegraphics[width=0.9\textwidth]{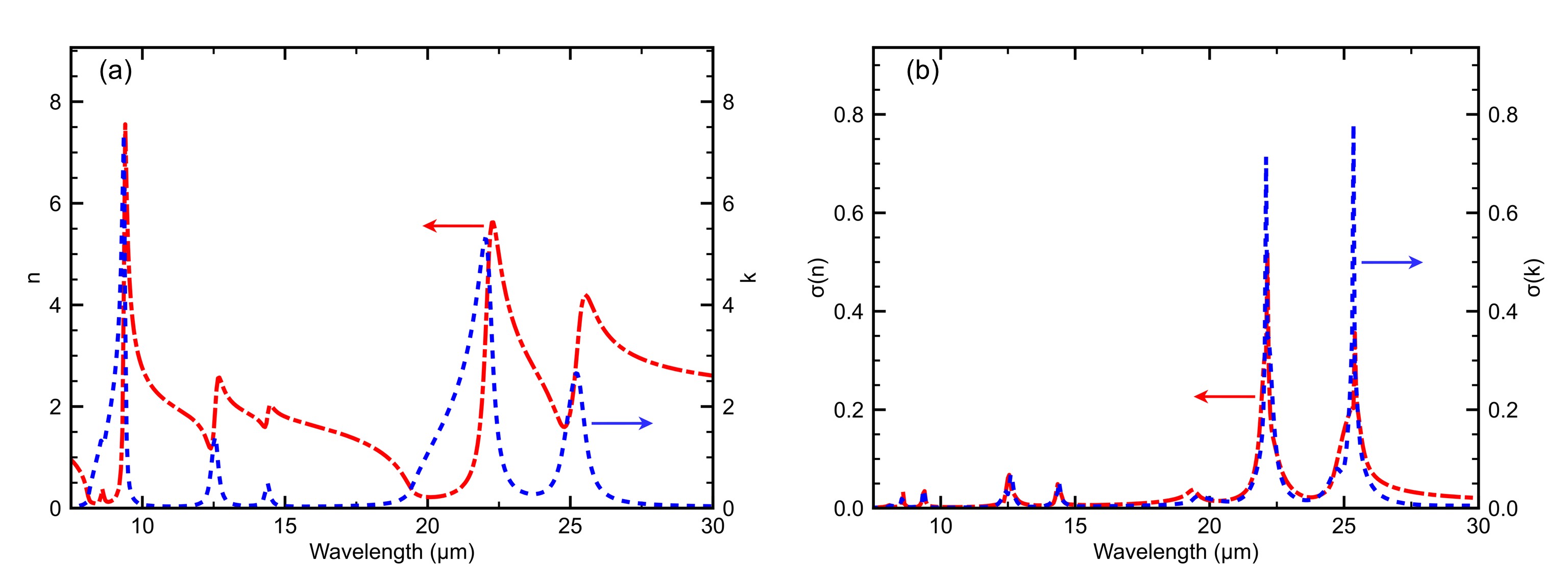}
    \caption[] {(a) Real (\textit{n}) and imaginary (\textit{k}) parts of the in-plane complex refractive index of quartz at room temperature, extracted from spectroscopic ellipsometry measurements. (b) Corresponding standard deviations of \textit{n} and \textit{k}, representing the measurement uncertainty associated with the ellipsometric fitting.}
  \label{fig:S5_PTR}
\end{figure}

\begin{figure}
 \includegraphics[width=0.9\textwidth]{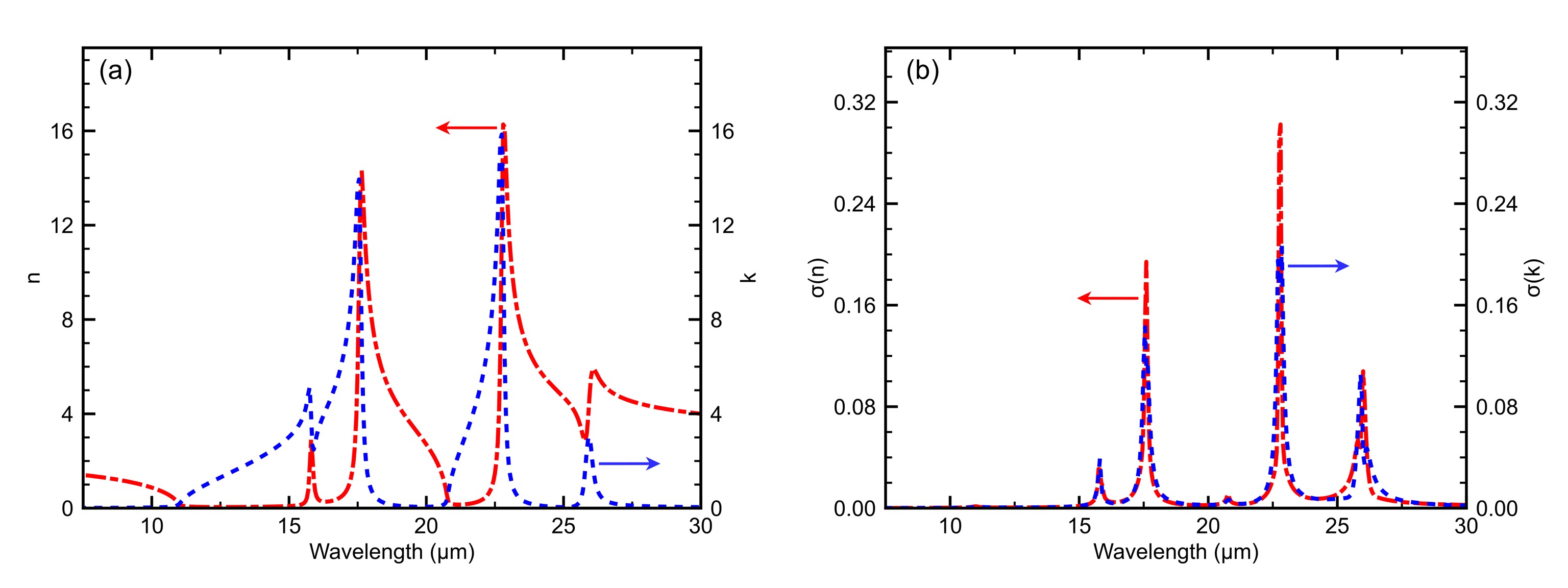}
    \caption[] {(a) Real (\textit{n}) and imaginary (\textit{k}) parts of the in-plane complex refractive index of sapphire at room temperature, extracted from spectroscopic ellipsometry measurements. (b) Corresponding standard deviations of \textit{n} and \textit{k}, representing the measurement uncertainty associated with the ellipsometric fitting.}
  \label{fig:S6_PTR}
\end{figure}

\begin{figure}
 \includegraphics[width=0.9\textwidth]{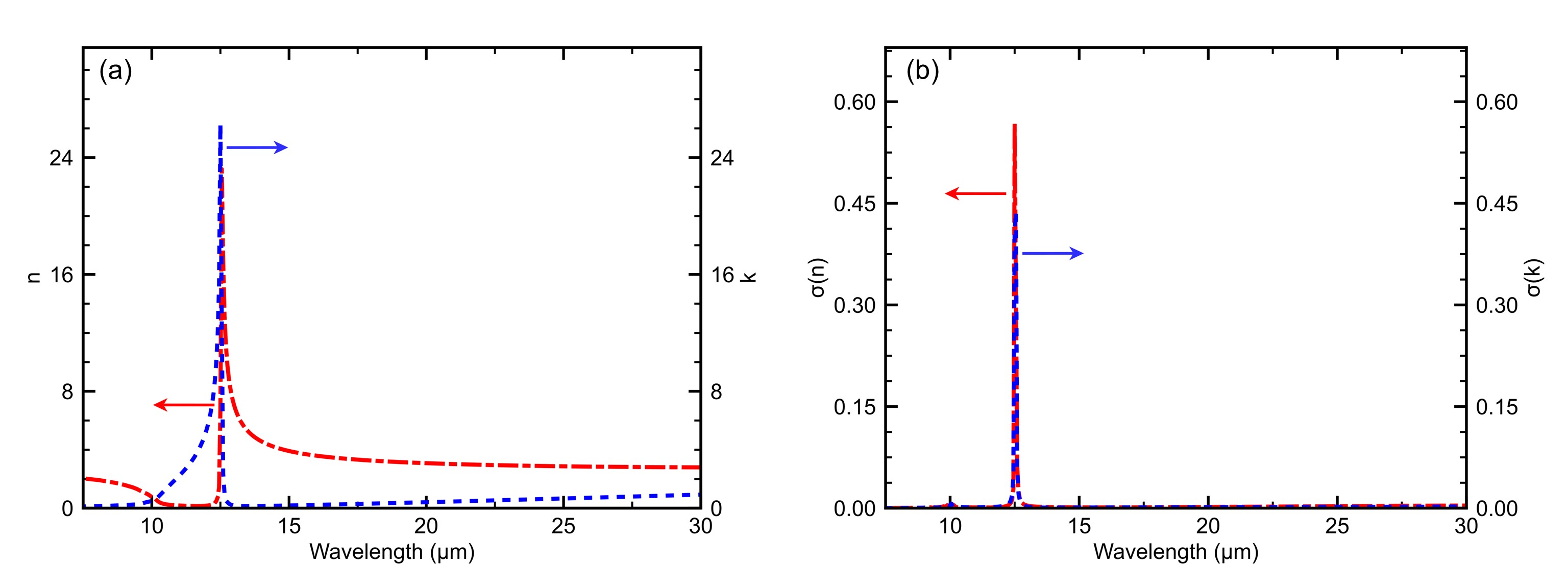}
    \caption[] {(a) Real (\textit{n}) and imaginary (\textit{k}) parts of the complex refractive index of SiC at room temperature, extracted from spectroscopic ellipsometry measurements. (b) Corresponding standard deviations of \textit{n} and \textit{k}, representing the measurement uncertainty associated with the ellipsometric fitting.}
  \label{fig:S7_PTR}
\end{figure}

\begin{figure}
 \includegraphics[width=0.9\textwidth]{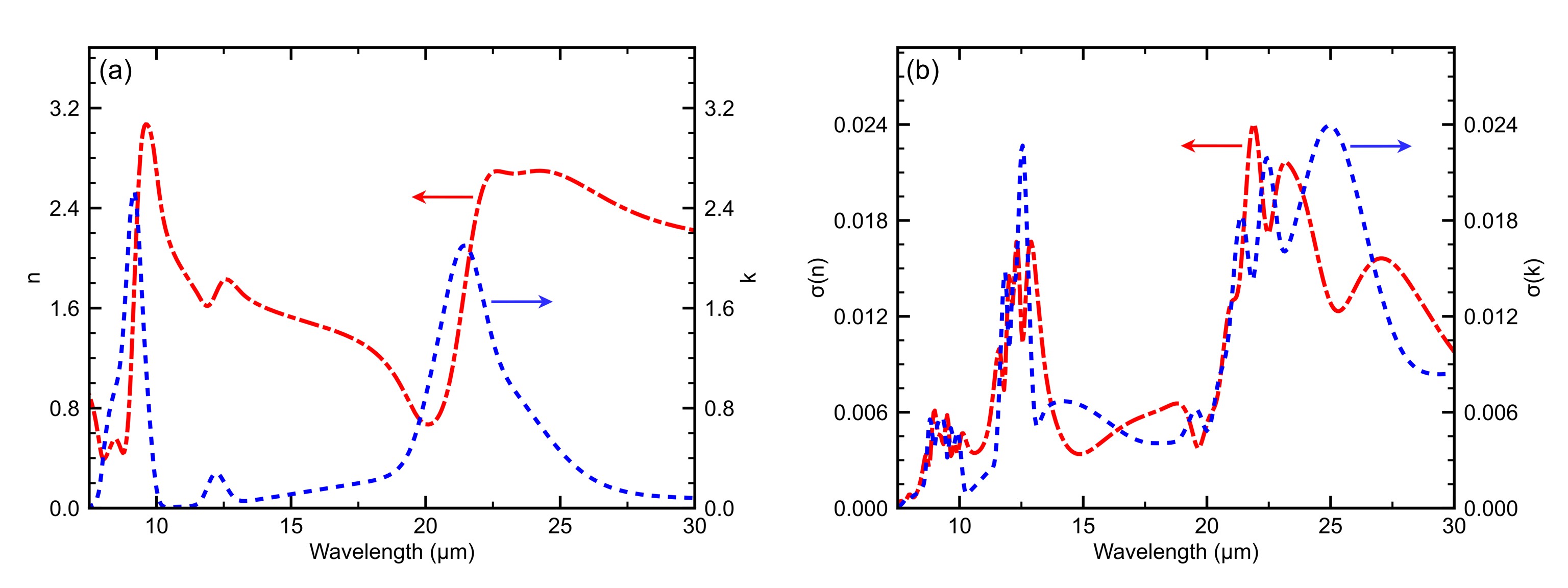}
    \caption[] {(a) Real (\textit{n}) and imaginary (\textit{k}) parts of the complex refractive index of thermal oxide at room temperature, extracted from spectroscopic ellipsometry measurements. (b) Corresponding standard deviations of \textit{n} and \textit{k}, representing the measurement uncertainty associated with the ellipsometric fitting.}
  \label{fig:S8_PTR}
\end{figure}

Using this validated fitting framework, the wavelength-dependent real ($n$) and imaginary ($k$) parts of the complex refractive index were extracted for all samples. Panels~(a) of Figs.~\ref{fig:S2_PTR}--\ref{fig:S8_PTR} present the extracted $n(\lambda)$ and $k(\lambda)$ spectra at 25~$^\circ$C. Panels~(b) of Figs.~\ref{fig:S2_PTR}--\ref{fig:S8_PTR} show the corresponding standard deviations of $n$ and $k$ at 25~$^\circ$C, which quantify the uncertainty associated with the ellipsometric fitting. Optical constants extracted at 200~$^\circ$C were used in the temperature-dependent analysis described in the main text but are not shown in Figs.~\ref{fig:S2_PTR}--\ref{fig:S8_PTR} for clarity.

Monte Carlo analysis was used to estimate the uncertainty in the extracted optical constants. For each dataset, WVASE generated synthetic $\Psi$ and $\Delta$ spectra by statistically resampling the measured ellipsometry data. Each resampled dataset was then refit using the same optical model and fitting procedure applied to the original measurements. This process produced distributions of fitted values $n(\lambda)$ and $k(\lambda)$ that reflect the sensitivity of the inversion to measurement noise and internal variability in the ellipsometric data. The uncertainty at each wavelength was defined as the standard deviation of the corresponding Monte Carlo distribution, providing a one-standard-deviation confidence interval for $n(\lambda)$ and $k(\lambda)$. The resulting optical constants and their associated uncertainties, shown in Figs.~\ref{fig:S2_PTR}--\ref{fig:S8_PTR}, provide a statistically grounded and internally consistent characterization of the optical response of all materials studied.

\section*{Section C: Penetration depth}

The penetration depth, $\delta_p$, at each wavelength is calculated from extracted extinction coefficient as
\begin{equation}
    {\delta_p} = \frac{\lambda}{4\pi k}
\end{equation}

The computed penetration depths for each dielectric material measured in this work at room temperature are illustrated in Fig. \ref{fig:S9_PTR}.

\section*{Section D: Transient Thermoreflectance Measurement and Two-Temperature Model}

\begin{figure}
 \includegraphics[width=0.9\textwidth]{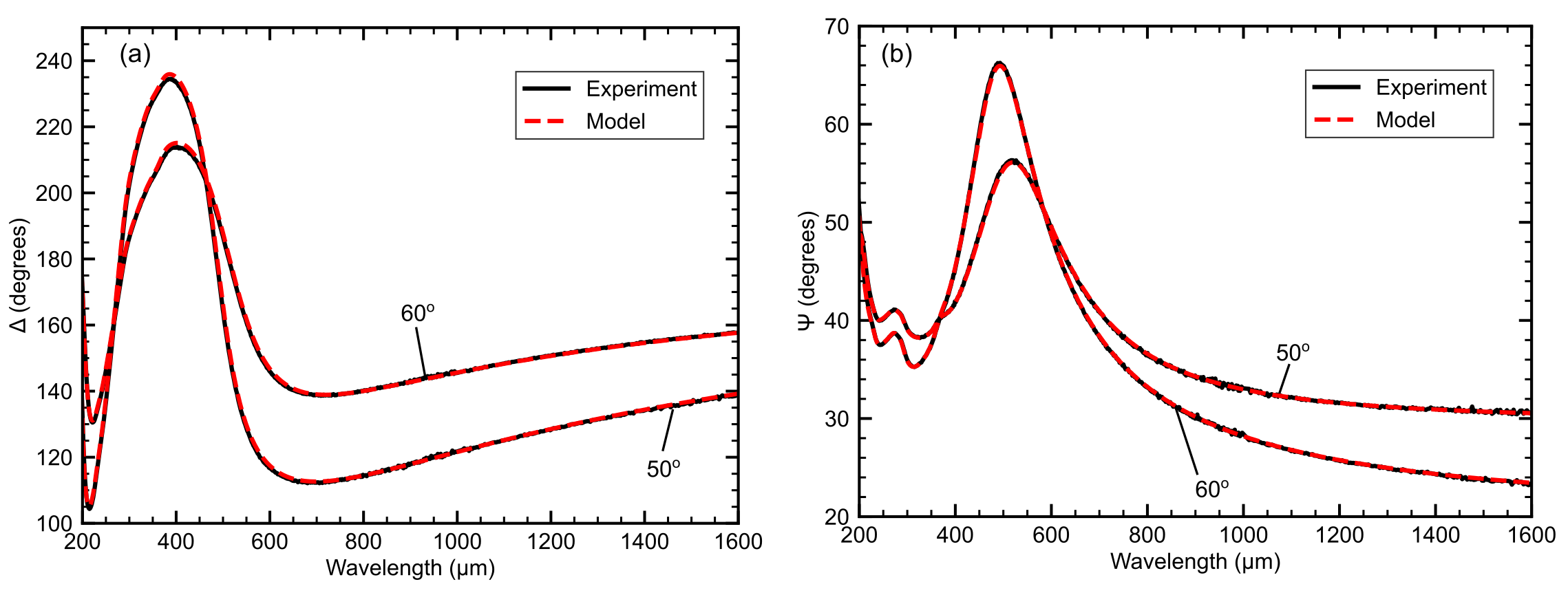}
    \caption[] {Visible ellipsometry data for thermally grown SiO$_2$ on Si, measured at incidence angles of 50\textdegree and 60\textdegree, are shown together with the corresponding model fits, from which a thermal oxide thickness of 103 nm was extracted.}
  \label{fig:SX_PTR}
\end{figure}

In the transient thermoreflectance measurement, schematically shown in Fig. \ref{fig:1_PTR}a, a 32W Spectra-Physics 1040 nm seed laser is split into two paths. The probe path utilizes an optical parametric amplifier (OPA) to tune the wavelength between 2–16 microns, while the pump path is directed through a second harmonic generation crystal to produce a 520 nm beam. The probe is delayed relative to the pump (via advancing the pump) and also guided to the sample at an off-angle to suppress optical cross-talk between the pump and probe beams at the detector. The reflected probe is spatially filtered from the pump and focused onto an MCT detector. Since the thermal oxide layer is transparent to the pump laser, the pump is absorbed in silicon and heats the oxide through the interfacial thermal boundary conductance. Visible-range spectroscopic ellipsometry, analyzed with an appropriate optical model, determined the thickness of the thermally grown SiO$_2$ layer on silicon to be 103 nm. Figure \ref{fig:SX_PTR} shows the corresponding measured ellipsometric data together with the model fit used to extract this thickness.

\begin{figure}
 \includegraphics[width=0.9\textwidth]{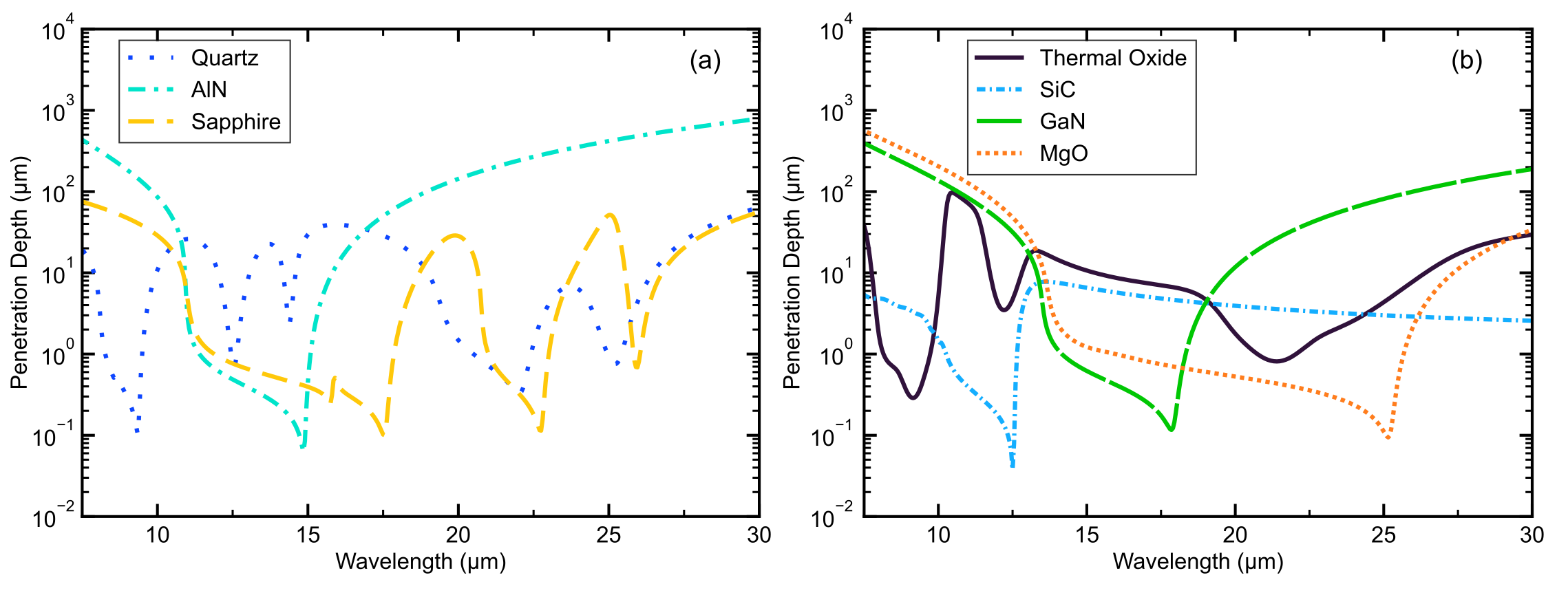}
    \caption[] {Penetration depth for the dielectrics measured in this study calculated from Eq. (S1) using the optical properties extracted from ellipsometry measurements at room temperature.}
  \label{fig:S9_PTR}
\end{figure}

To fit for the thermal boundary conductance (TBC) between the silicon and thermal oxide, $h_{\mathrm{SiO_{2}/Si}}$, we utilize a modified two-temperature model described by the following equations:

\begin{equation}
    C_{\mathrm{SiO_{2}}}\frac{\partial T_{\mathrm{SiO_{2}}}}{\partial t} = \nabla(k_{\mathrm{SiO_{2}}}\nabla T_{\mathrm{SiO_{2}}}) |x=0
\end{equation}

\begin{equation}
C_{e}\frac{\partial T_{e}}{\partial t} = \nabla(k_{e}\nabla T_{e}) -G(T_e -T_l) +S(x,t) 
|x \ne 0
\end{equation}

\begin{equation}
C_{l}\frac{\partial T_{l}}{\partial t} = \nabla(k_{l}\nabla T_{l}) +G(T_e -T_l) |x \ne 0
\end{equation}

\noindent where $C$ describes the heat capacity, $k$ is the thermal conductivity of the respective channel, $G$ represents the electron-phonon coupling factor, and $S(x,t)$ describes the source term utilized in our thermal model. The source term is expressed by the following equation: 
\begin{equation}
S(x,t) = (1-R)\frac{1.76J}{t_p}\mathrm{sech}^2(\frac{1.76t}{t_p})\frac{dI}{dx}
\end{equation}
where $R$ is the reflectance, $J$ is the incident fluence, $t_p$ is the pulse width of the laser, and ${dI}/{dx}$ is the spatial derivative of the light intensity profile determined from the transfer matrix method with optical constants measured at the pump wavelength of 520 nm. We measure the complex refractive indices of the Si and SiO$_2$ at 520 nm to be $4.2 + .029 i$ and $1.46$, respectively. 
The interface between the SiO$_2$ and Si is defined as a 0.5 nm node with a thermal conductivity that can be expressed as $h_{\mathrm{Si/SiO_{2}}} \times d_{\mathrm{interface}}$ and a heat capacity of zero. Equation (S2) describes the temperature in the oxide layer, where we neglect the electron contribution within this channel due to it being highly insulating. Equations (S3) and (S4) describe the coupled equations for the electronic and phononic temperature in the underlying silicon substrate. We use literature values for the silicon electron-phonon coupling factor and phonon thermal properties \cite{medvedevElectronPhononCouplingSemiconductors2023,braunSteadystateThermoreflectanceMethod2019}. We adjust the electronic heat capacity to match the electron-phonon coupling time constant, $\tau_{e-p}=C_e/G$, of silicon with the onset of the oxide response in our data around 300 ps which results in $C_e = 3 \times 10^5 ~\mathrm{J~m^{-3}~K^{-1}}$. As the re-rise in our thermoreflectance magnitude is directly linked to the increasing temperature of the oxide layer, we fit the temperature of this layer to the re-rise fitting for TBC between the SiO$_2$ and silicon. Since the penetration depth of our infrared probe is larger than the thickness of the SiO$_{2}$ film, as seen in Fig. \ref{fig:S9_PTR}, we use the average temperature of the silica as our calculated thermoreflectance. By creating a curve of the fitting residual as a function of the thermal boundary conductance, we can analyze the sensitivity of the model to variations in the TBC. We define the fitting residual as a residual percent, $Z_{\%}$, following Feser \textit{et al.} \cite{feserProbingAnisotropicHeat2012}:

\begin{equation}
    {Z_{\%}} = \left|\frac{R(h_{\mathrm{exact}})-R(h_{\mathrm{perturbed}}) }{R(h_{\mathrm{exact}})}\right| * 100
\end{equation}

\begin{figure}
 \includegraphics[width=0.7\textwidth]{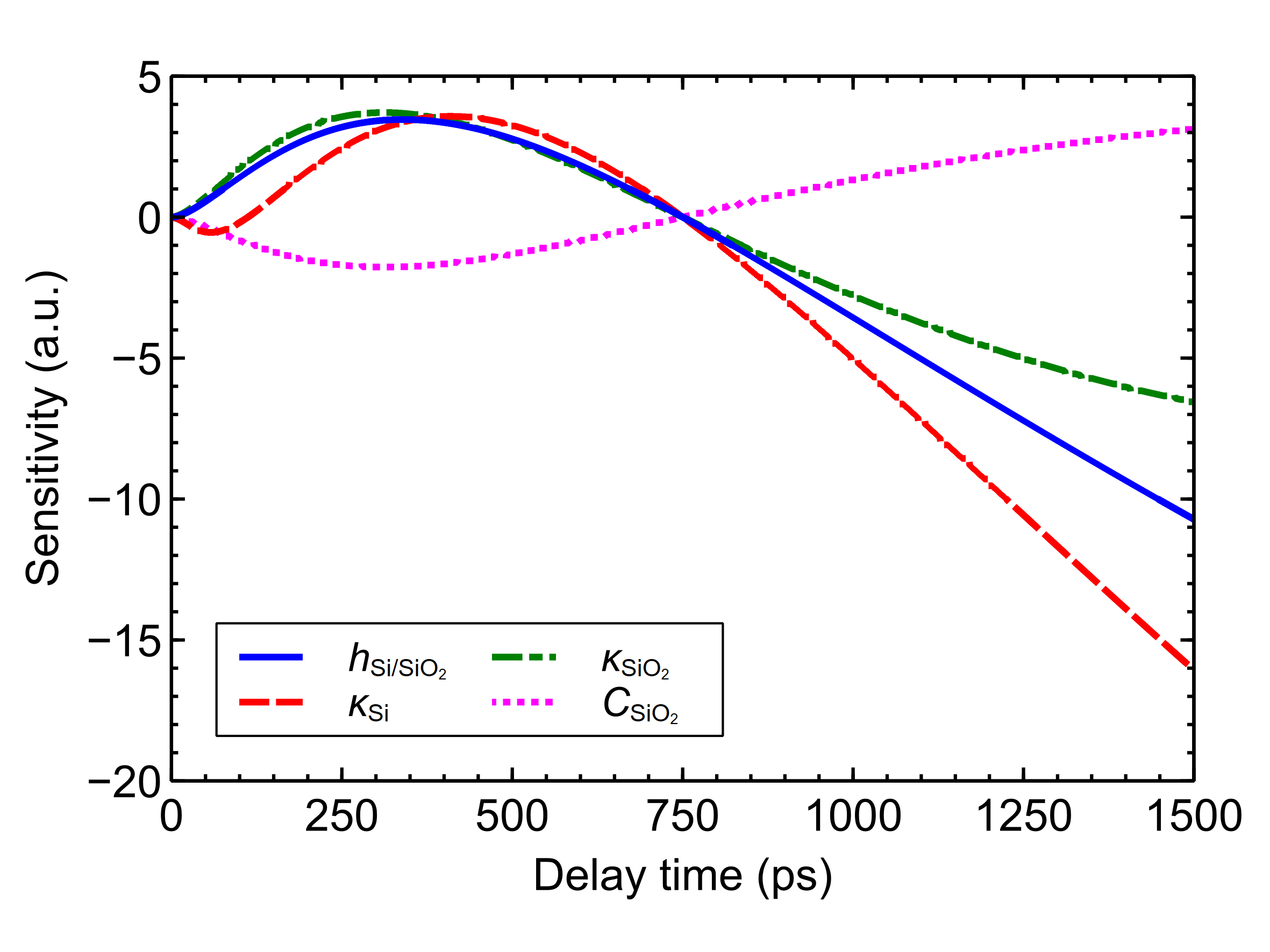}
    \caption[] {Two-temperature model sensitivities to thermal parameters calculated using a TBC value of 160~MW\,m$^{-2}$\,K$^{-1}$. We note a strong sensitivity to the thermal boundary conductance and the silicon thermal conductivity. The sensitivity to the thermal boundary conductance is further enhanced at lower TBC values allowing us to set a lower bound.}
  \label{fig:S10_PTR}
\end{figure}

The inset in Fig. \ref{fig:1_PTR}c shows the normalized fitting residual as a function of the thermal boundary conductance between silicon and SiO$_2$, obtained from fitting the thermoreflectance data using the modified two-temperature model. The residual exhibits a smooth and well-defined minimum of approximately 6\% at a TBC near 160~MW\,m$^{-2}$\,K$^{-1}$, identifying the interfacial conductance value that best reproduces the experimental response. The curvature of the residual around this minimum reflects the sensitivity of the fit to variations in the assumed TBC. The residual increases only weakly for larger TBC values, indicating limited sensitivity to the upper bound. As a result, while the analysis supports a lower bound of approximately 160~MW\,m$^{-2}$\,K$^{-1}$ for the SiO$_2$/Si interfacial conductance, it does not justify assignment of a definitive upper limit under the present modeling framework. The sensitivity of our model to the thermal parameters is also shown in Fig. \ref{fig:S10_PTR} for a thermal boundary conductance of 160 MW\,m$^{-2}$\,K$^{-1}$. Here, we see a significant sensitivity to the thermal boundary conductance along with the silicon thermal conductivity. Importantly, this sensitivity to the boundary conductance becomes the predominant sensitivity within our thermal model at lower thermal boundary conductance values, allowing us to accurately prescribe a lower bound for the TBC.

\section*{Section E: Calculated Figure of Merits and Interpretations}

The figure of merit (FOM) introduced in the main text is defined as
\begin{equation}
    FOM = \alpha(\lambda_{pump}) \left| \frac{dR}{dT}(\lambda_{probe})\right|
\end{equation}
where $\alpha$ is the absorptivity at the pump wavelength and $dR/dT$ is the thermoreflectance coefficient at the probe wavelength. This metric is intended as a comparative screening tool rather than as a predictor of absolute signal strength in practical thermoreflectance measurements.

In time-domain thermoreflectance (TDTR) experiments, the detected signal depends on the temperature rise induced by the pump beam, the thermoreflectance coefficient $dR/dT$, and the probe beam intensity. Under fixed incident pump fluence, the absorbed power scales with $\alpha$, such that the FOM captures the combined influence of pump absorption and optical readout sensitivity. However, in practical measurements, the experimental constraint is typically a maximum allowable temperature rise, imposed to maintain linear response and avoid sample damage. Under such fixed temperature-rise conditions, the detected signal amplitude is governed primarily by $dR/dT$ and the probe intensity, while absorptivity determines the pump fluence required to reach that temperature rise.

Accordingly, the FOM should not be interpreted as an optimization metric for maximizing heating efficiency or signal amplitude. Instead, it provides a convenient means to identify wavelength regimes where strong thermoreflectance sensitivity can be achieved without requiring excessively high pump fluence. This interpretation is consistent with the emphasis in the main text that $dR/dT$ is the dominant performance metric for thermoreflectance transducers.

Regarding the form of the FOM, while $\alpha$ is bounded between 0 and 1, the thermoreflectance coefficient $dR/dT$ does not possess a strict mathematical upper bound. In practice, however, $dR/dT$ is physically constrained by optical dispersion relations, phonon linewidths, and temperature-induced broadening of resonances. These effects limit the achievable magnitude of $dR/dT$, ensuring that the FOM remains bounded for real materials and experimentally accessible conditions.

In evaluating the FOM, a transducer thickness of 80~nm is assumed for all materials. This thickness is representative of commonly used thermoreflectance transducers and allows direct comparison with previously reported figures of merit for metallic transducers~\cite{wilsonThermoreflectanceMetalTransducers2012b}. For the optically thin dielectric films considered here, variations in thickness primarily rescale absorptivity without altering the underlying wavelength dependence of $dR/dT$, and the qualitative trends reported in this work are therefore robust with respect to moderate thickness variations.

Full two-dimensional FOM maps as a function of pump and probe wavelength for all materials studied are provided in Figs. \ref{fig:S11_PTR}-\ref{fig:S21_PTR} to contextualize the wavelength-dependent thermoreflectance behavior discussed in the main text.


\newpage

\mbox{}

\newpage


\twocolumngrid

\noindent
\vspace{-15pt}
\includegraphics[width=0.95\columnwidth]{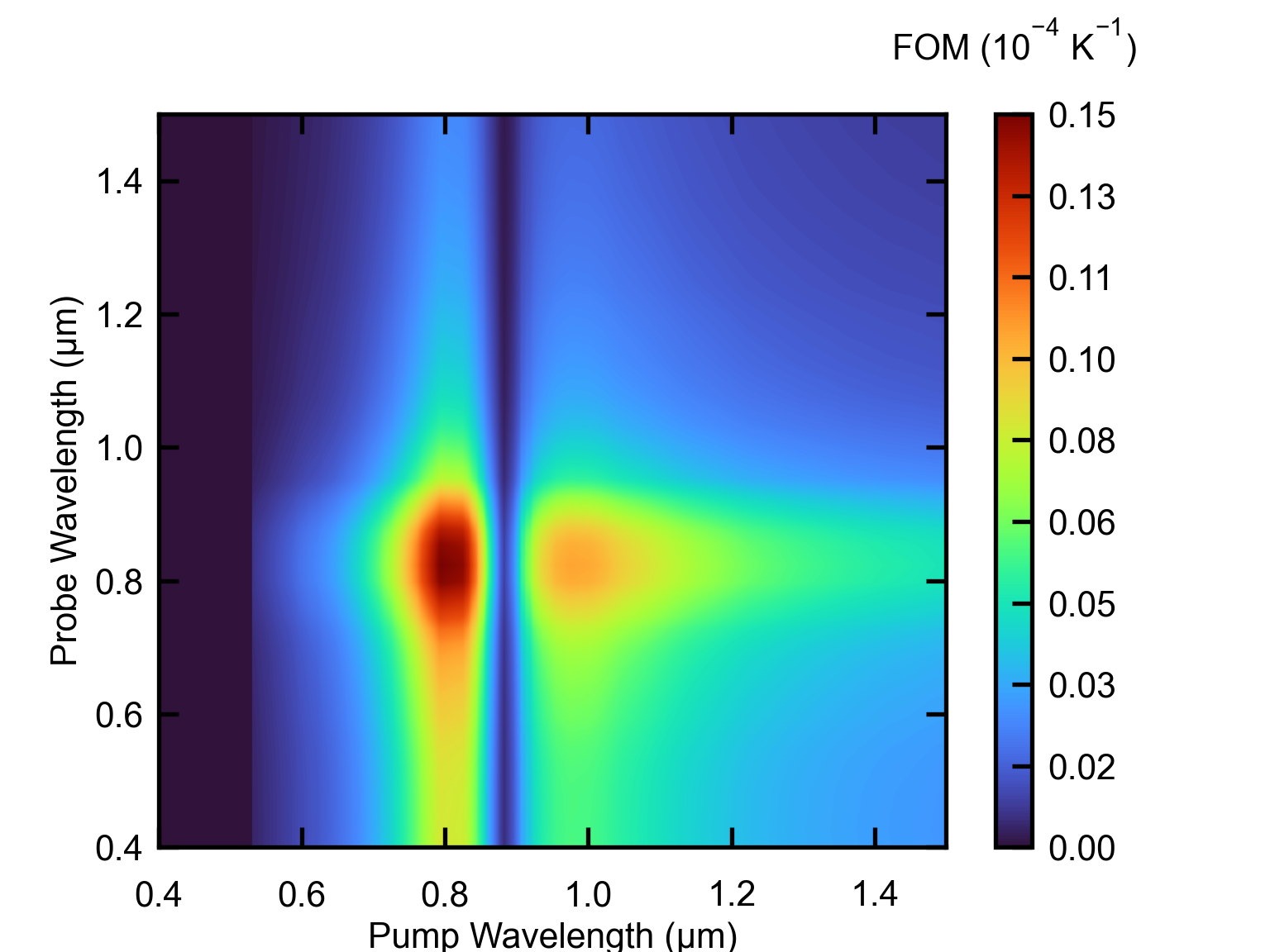}
\begin{figure}[h!]
\vspace{-10pt}
\caption[] {Computed FOM for Al. The thermoreflectance and extinction coefficient data were taken from Refs.~\cite{wilsonThermoreflectanceMetalTransducers2012b, palikHandbookOpticalConstants1998}. }
  \label{fig:S11_PTR}
\end{figure}

\noindent
\vspace{-15pt}
\includegraphics[width=0.95\columnwidth]{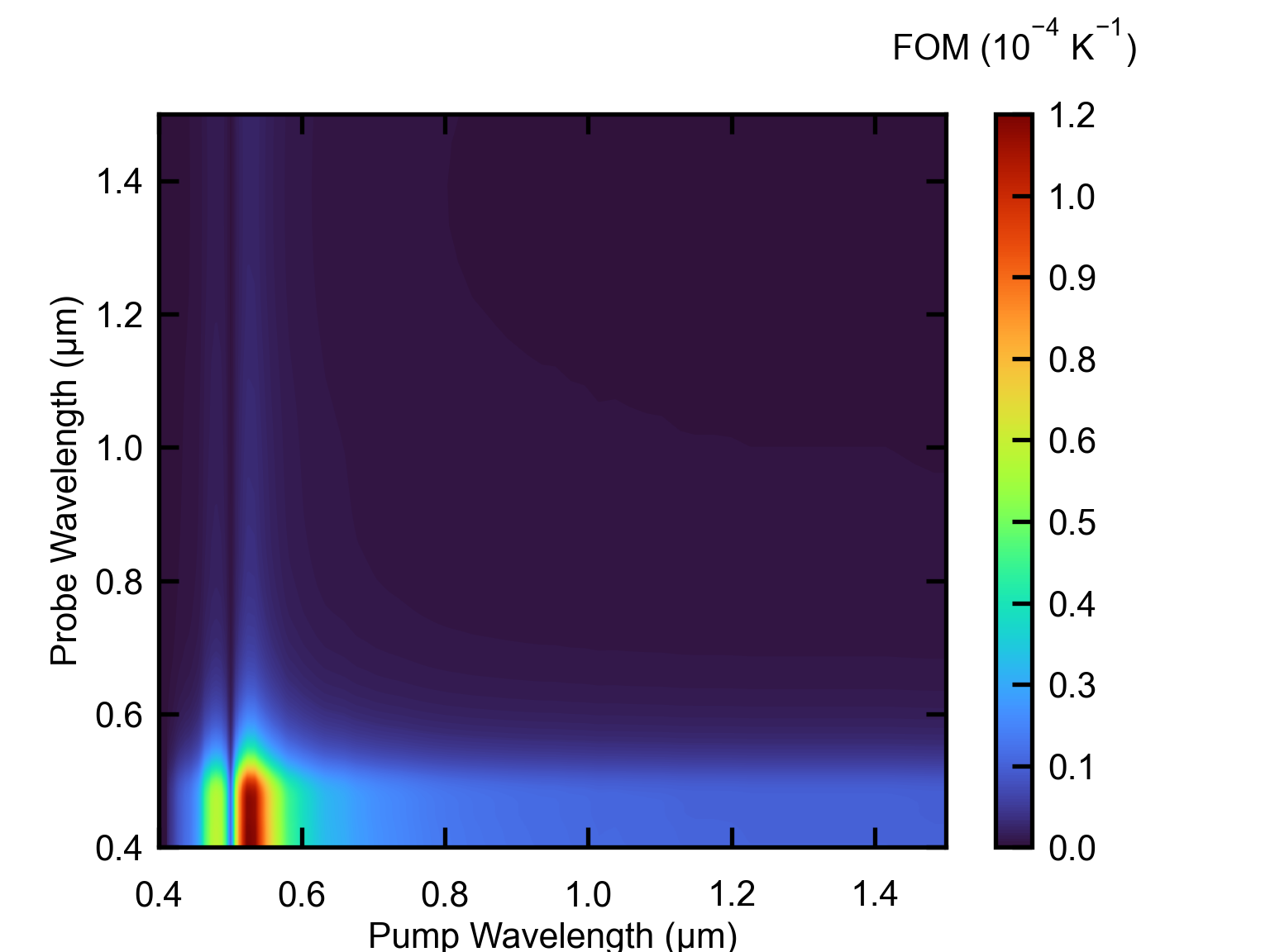}
\begin{figure}[h!]
\vspace{-10pt}
\caption[] {Computed FOM for Au. The thermoreflectance and extinction coefficient data were taken from Refs.~\cite{wilsonThermoreflectanceMetalTransducers2012b, palikHandbookOpticalConstants1998}.}
  \label{fig:S12_PTR}
\end{figure}

\noindent
\vspace{-15pt}
\includegraphics[width=0.95\columnwidth]{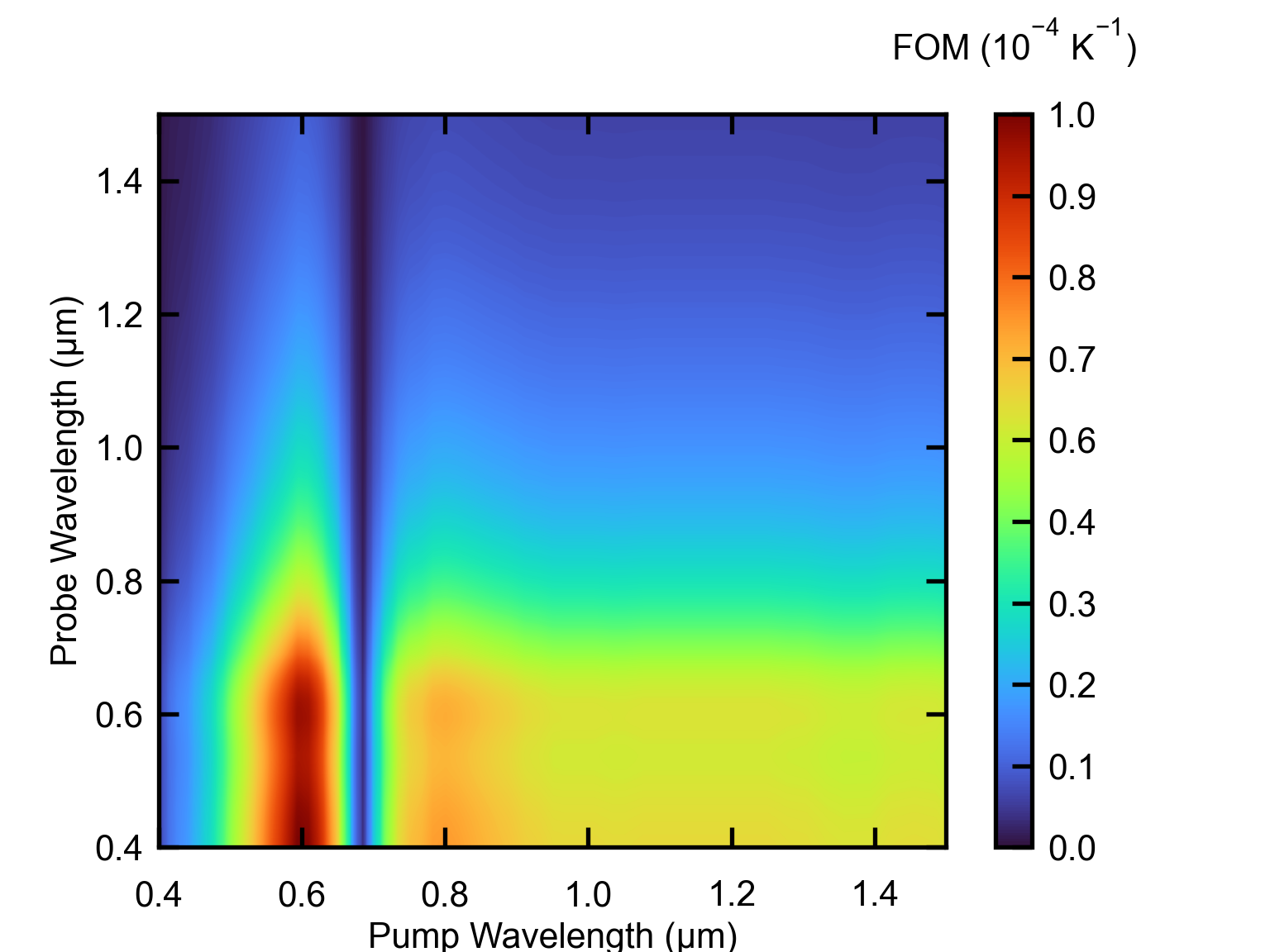}
\begin{figure}[h!]
\vspace{-10pt}
\caption[] {Computed FOM for Ta. The thermoreflectance and extinction coefficient data were taken from Refs.~\cite{wilsonThermoreflectanceMetalTransducers2012b, palikHandbookOpticalConstants1998}.}
  \label{fig:S13_PTR}
\end{figure}

\noindent
\vspace{-15pt}
\includegraphics[width=0.95\columnwidth]{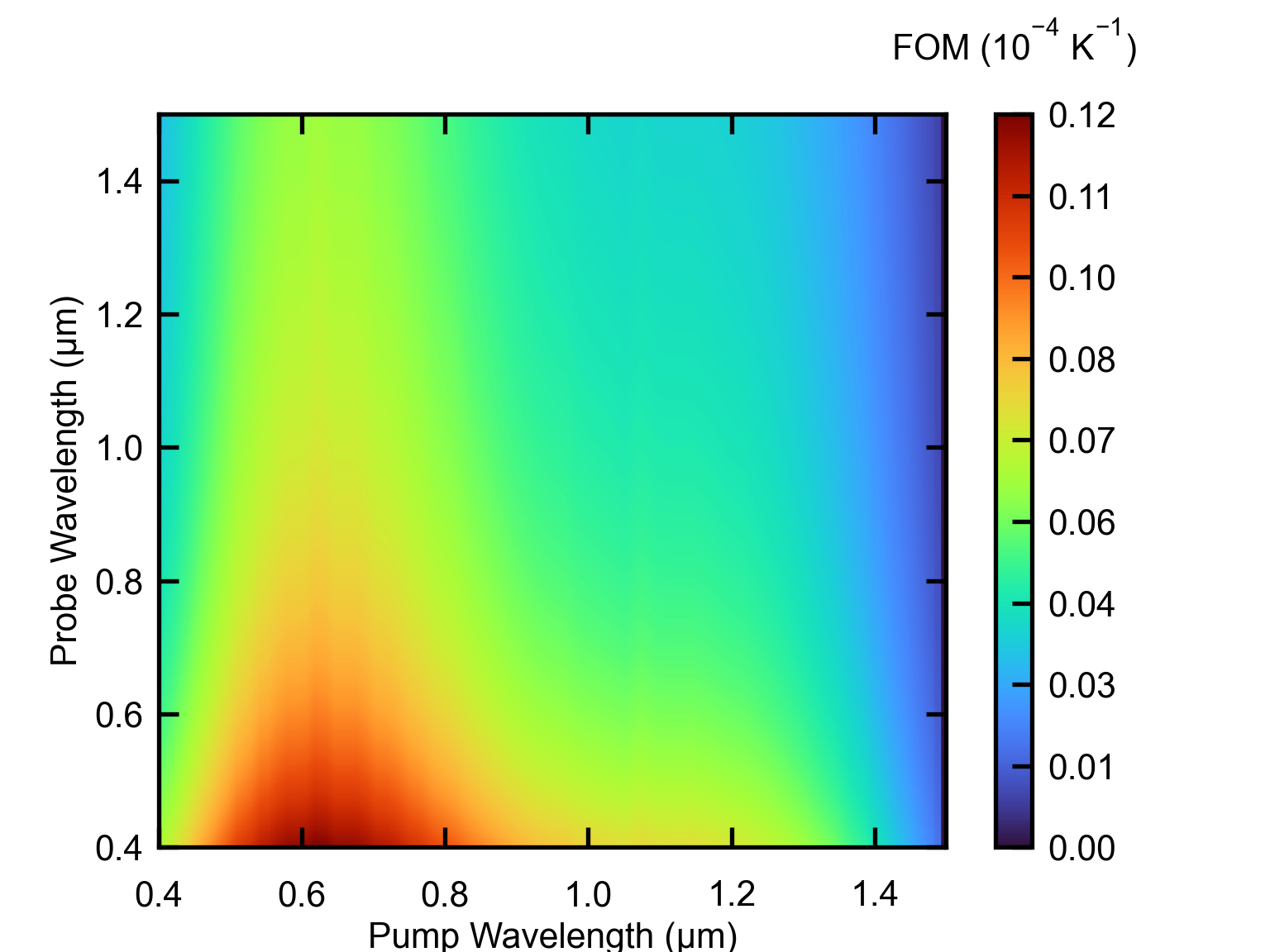}
\begin{figure}[h!]
\vspace{-10pt}
\caption[] {Computed FOM for Pt. The thermoreflectance and extinction coefficient data were taken from Refs.~\cite{wilsonThermoreflectanceMetalTransducers2012b, palikHandbookOpticalConstants1998}.}
  \label{fig:S14_PTR}
\end{figure}

\noindent
\vspace{-15pt}
\includegraphics[width=0.95\columnwidth]{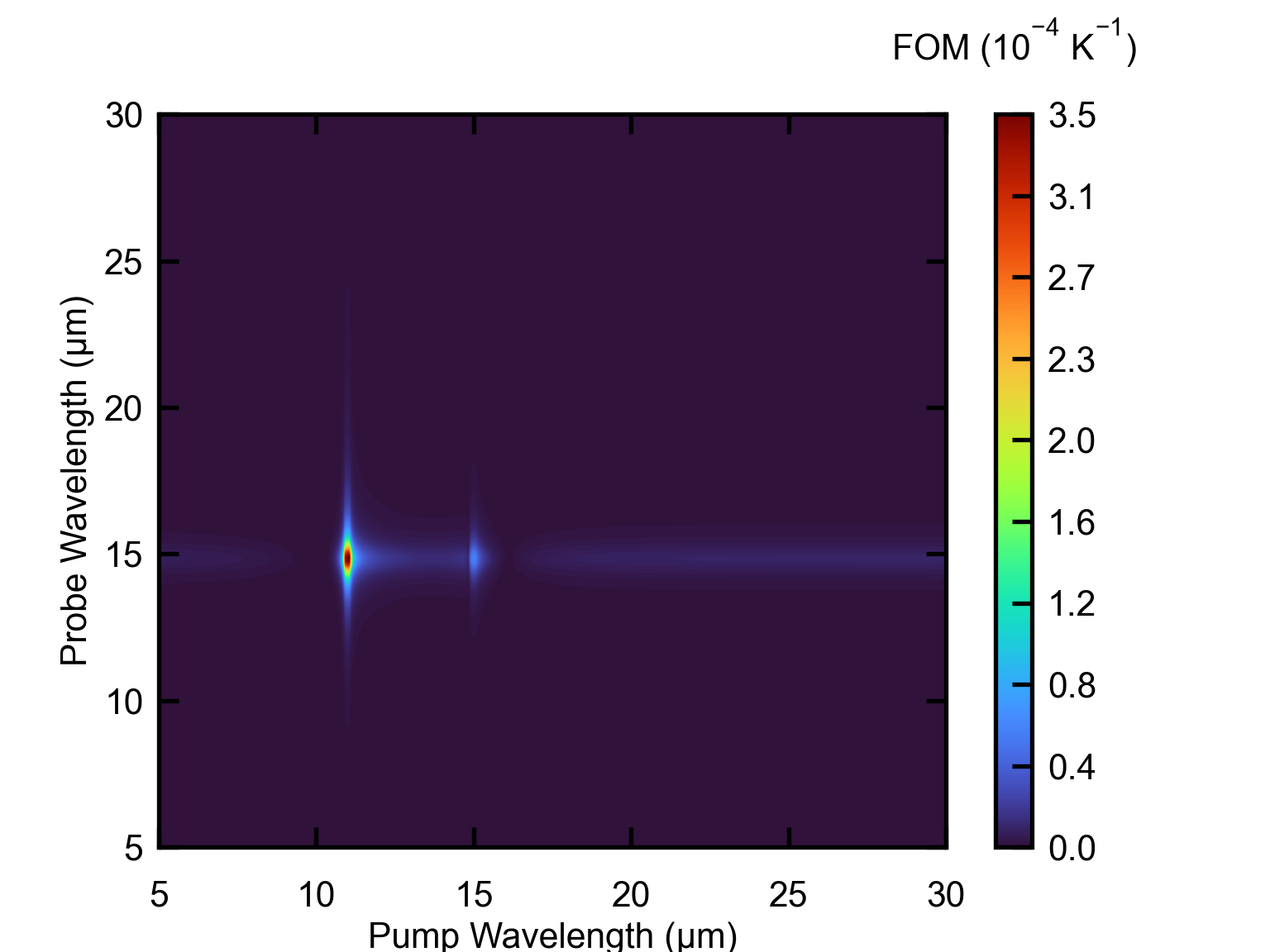}
\begin{figure}[h!]
\vspace{-10pt}
\caption[] {Computed FOM for AlN. The thermoreflectance and absorption data were measured in this work.}
  \label{fig:S15_PTR}
\end{figure}

\noindent
\vspace{-15pt}
\includegraphics[width=0.95\columnwidth]{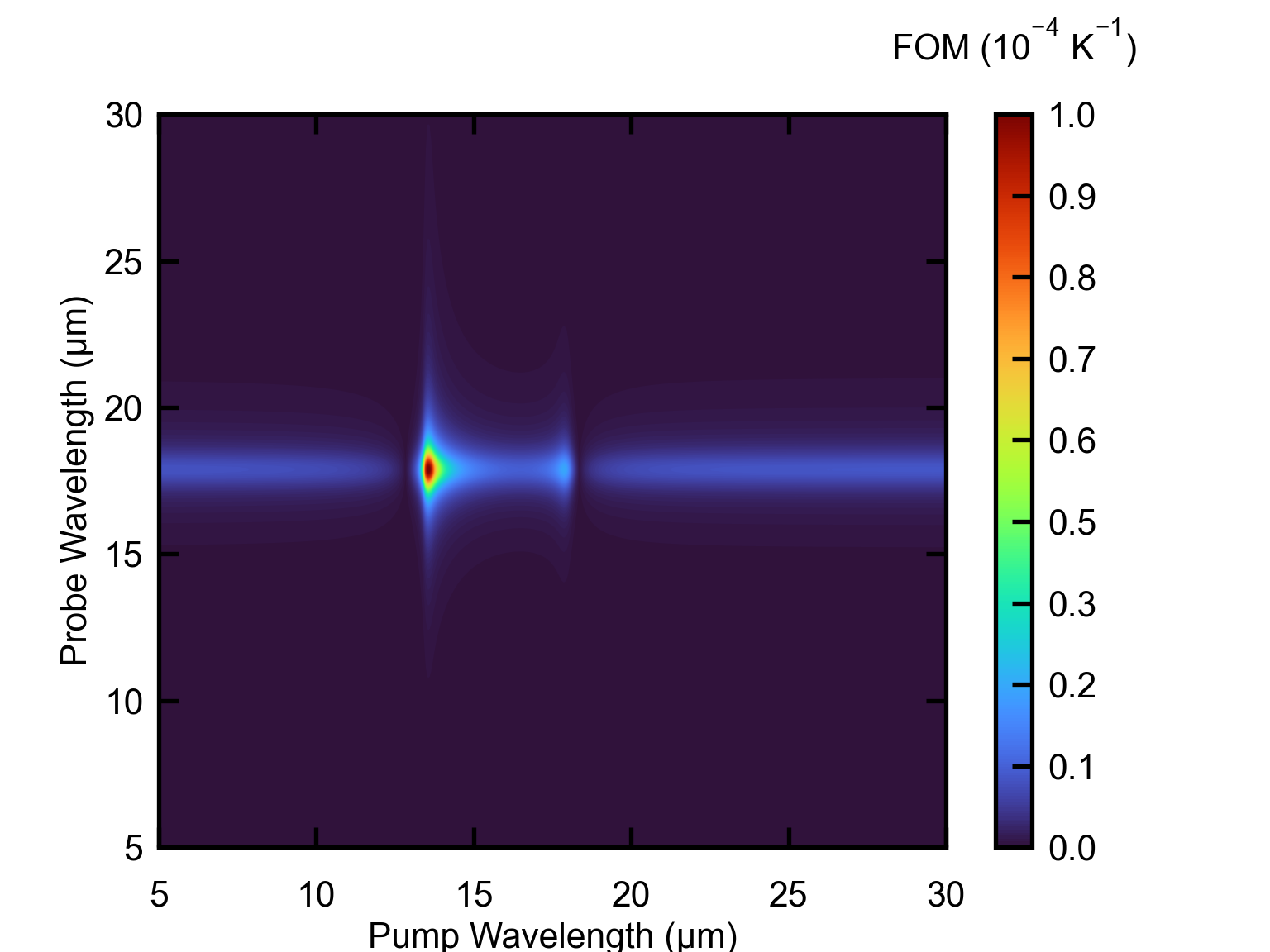}
\begin{figure}[h!]
\vspace{-10pt}
\caption[] {Computed FOM for GaN. The thermoreflectance and absorption data were measured in this work.}
  \label{fig:S16_PTR}
\end{figure}

\noindent
\vspace{-15pt}
\includegraphics[width=0.95\columnwidth]{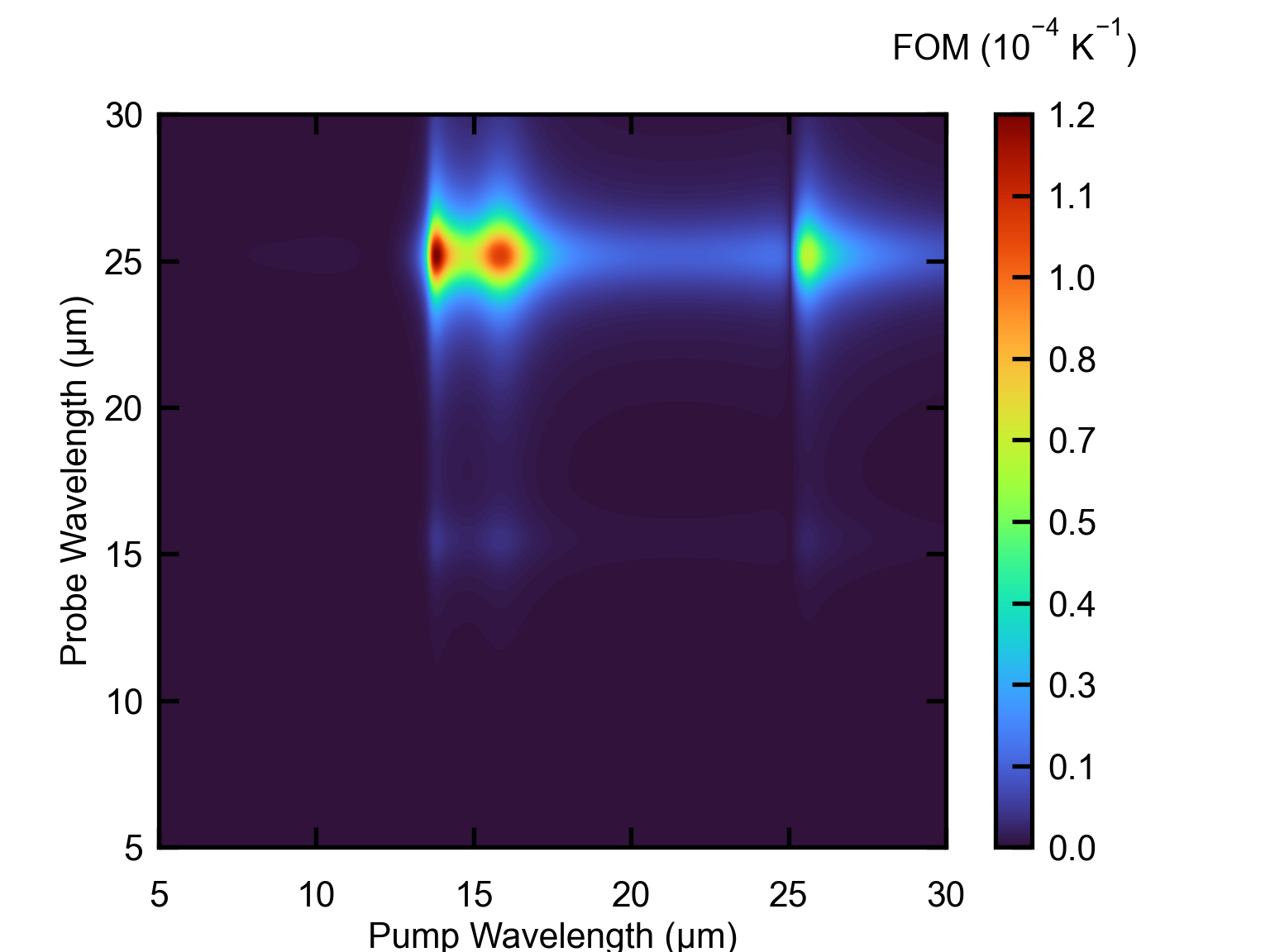}
\begin{figure}[h!]
\vspace{-10pt}
\caption[] {Computed FOM for MgO. The thermoreflectance and absorption data were measured in this work.}
  \label{fig:S17_PTR}
\end{figure}

\noindent
\vspace{-15pt}
\includegraphics[width=0.95\columnwidth]{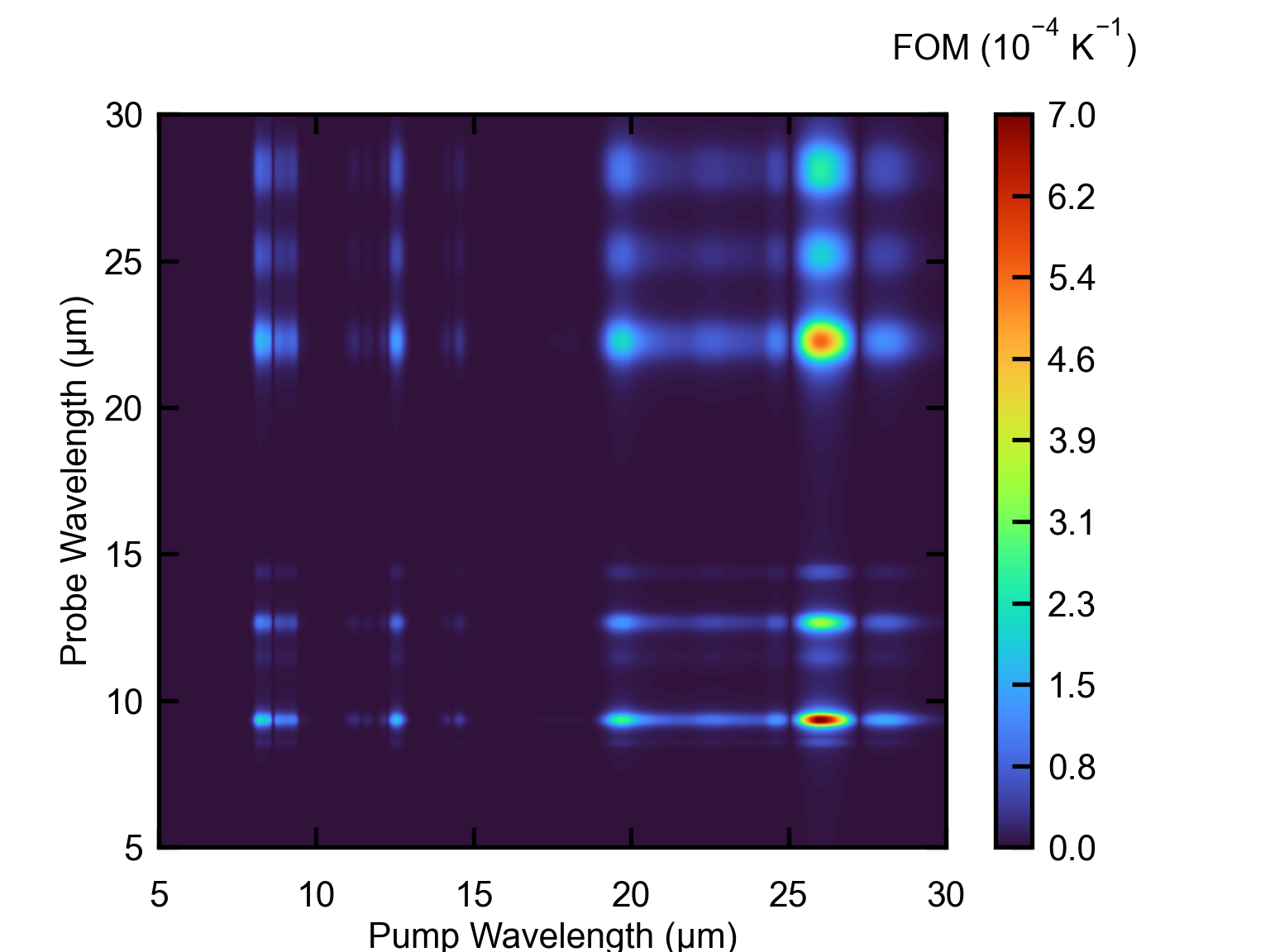}
\begin{figure}[h!]
\vspace{-10pt}
\caption[] {Computed FOM for quartz. The thermoreflectance and absorption data were measured in this work.}
  \label{fig:S18_PTR}
\end{figure}

\noindent
\vspace{-15pt}
\includegraphics[width=0.95\columnwidth]{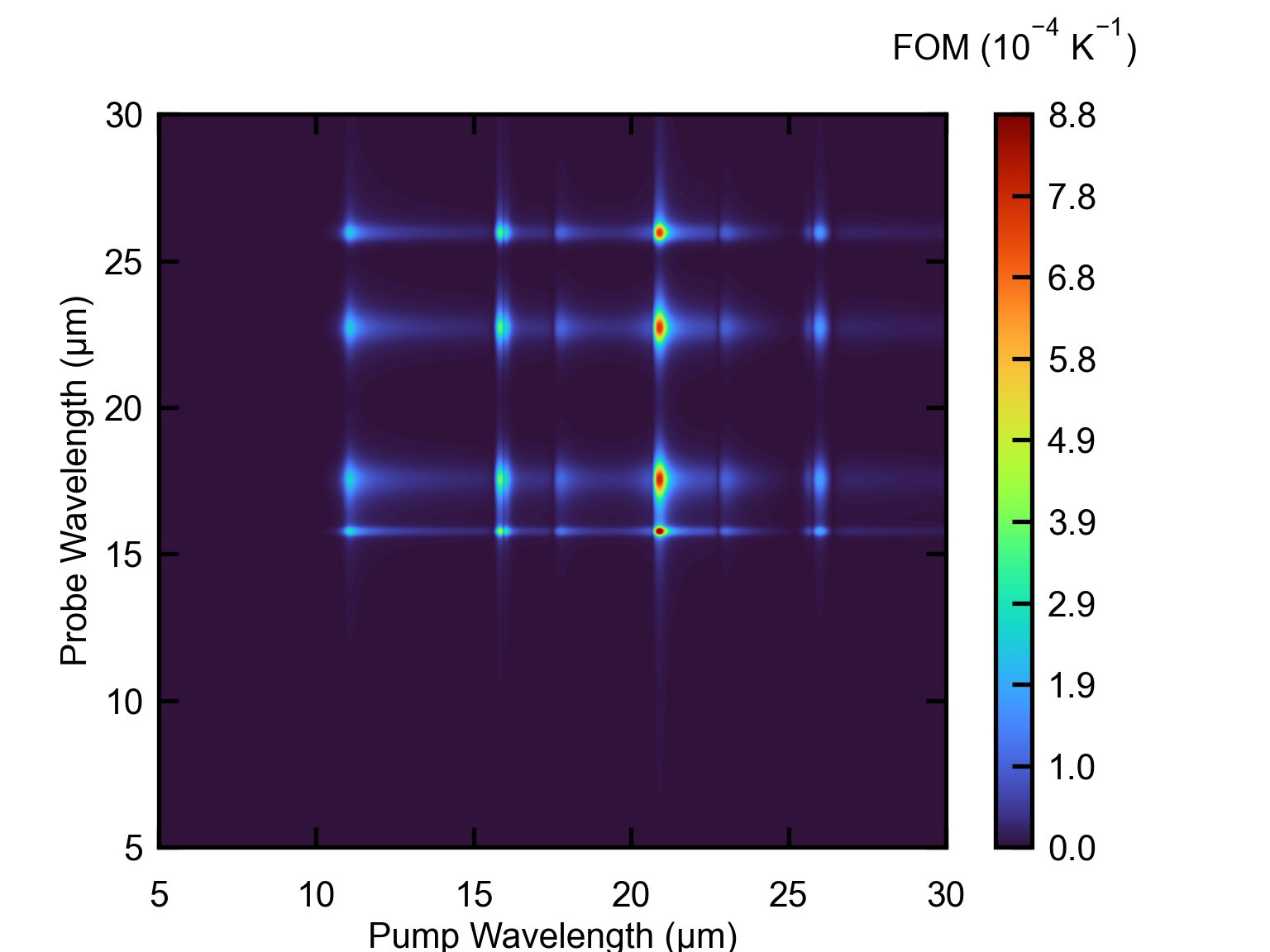}
\begin{figure}[h!]
\vspace{-10pt}
\caption[] {Computed FOM for sapphire. The thermoreflectance and absorption data were measured in this work.}
  \label{fig:S19_PTR}
\end{figure}

\noindent
\vspace{-1pt}
\includegraphics[width=0.95\columnwidth]{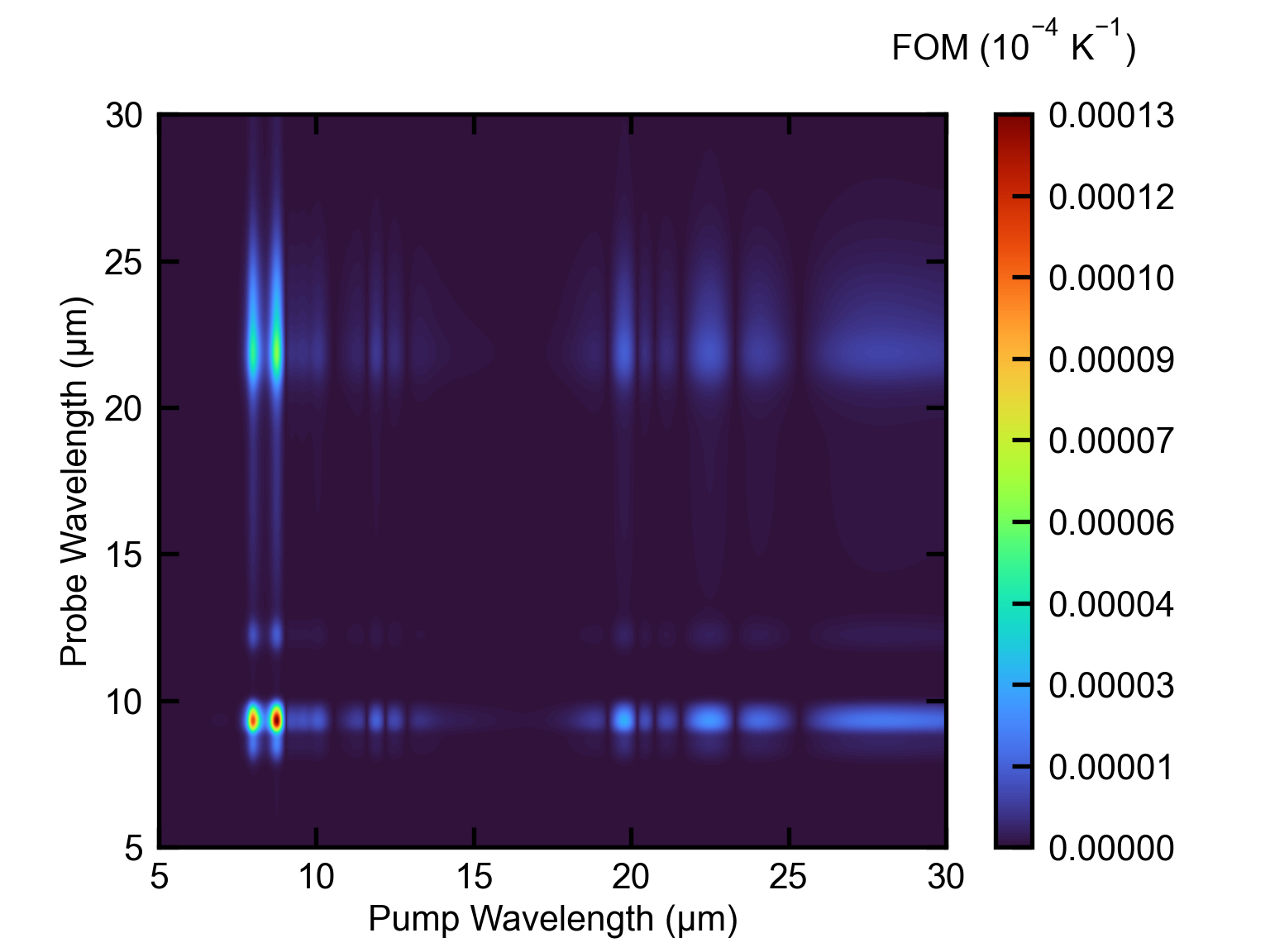}
\begin{figure}[h!]
\vspace{-10pt}
\caption[] {Computed FOM for thermal oxide. The thermoreflectance and absorption data were measured in this work.}
  \label{fig:S20_PTR}
\end{figure}

\vspace{13pt}

\noindent
\vspace{-1pt}
\includegraphics[width=0.95\columnwidth]{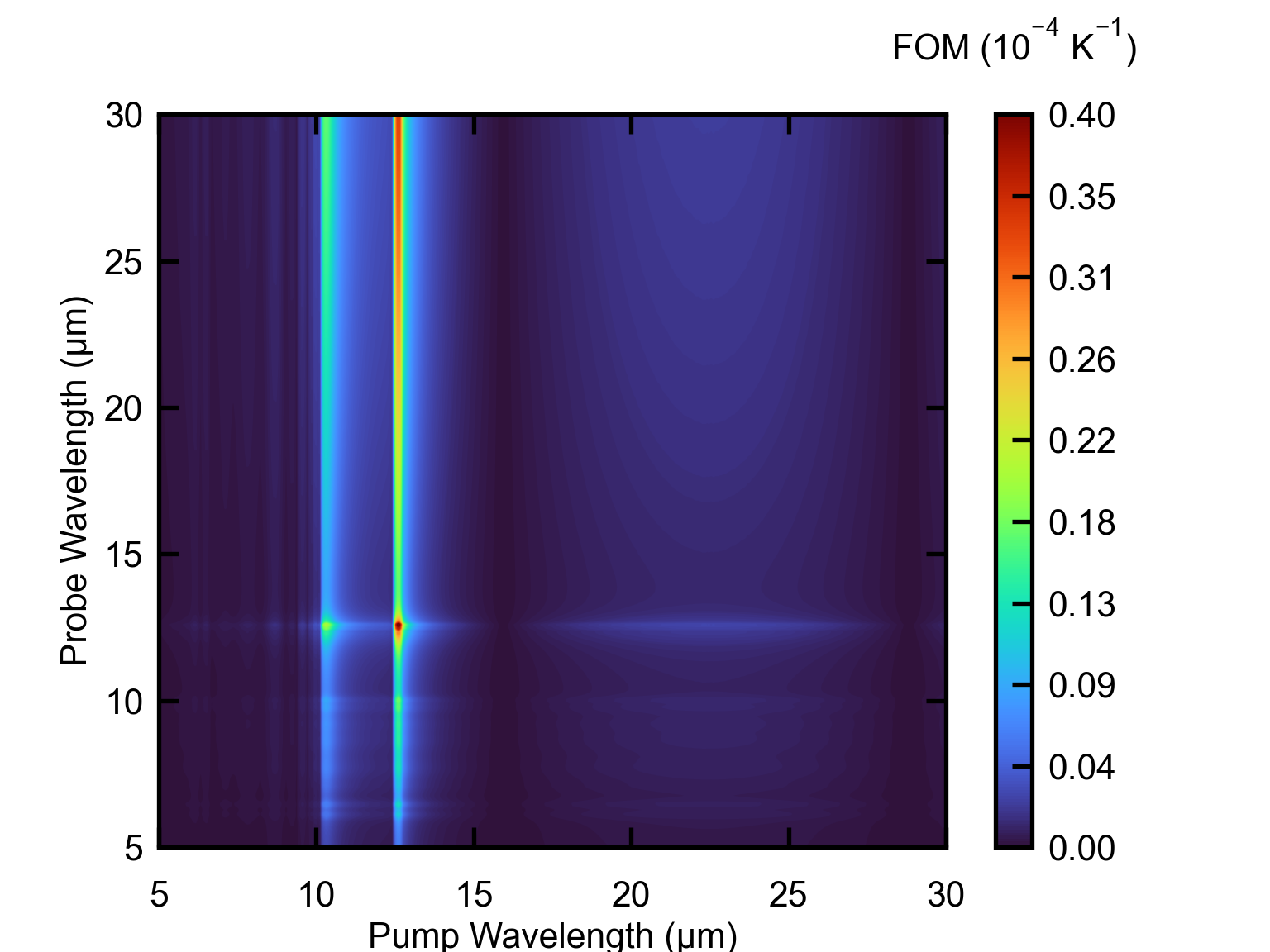}
\begin{figure}[h!]
\vspace{-10pt}
\caption[] {Computed FOM for SiC. The thermoreflectance and absorption data were measured in this work.}
  \label{fig:S21_PTR}
\end{figure}


\newpage


\end{document}